\documentclass{article}

\usepackage{arxiv}

\usepackage[utf8]{inputenc} 
\usepackage[T1]{fontenc}    
\usepackage[colorlinks]{hyperref}       
\usepackage{url}            
\usepackage{booktabs}       
\usepackage{amsfonts}       
\usepackage{nicefrac}       
\usepackage{color,xcolor}
\usepackage{microtype}      
\usepackage{graphicx}
\usepackage{amssymb}
\usepackage{amsmath}
\usepackage{subcaption}
\usepackage[numbers,sort&compress]{natbib}

\hypersetup{linkbordercolor=black, linkcolor = black}

\title{How do classroom-turnover times depend on lecture-hall size?}

\author{
  Joseph Benson\\
  Mathematics, Statistics, and Computer Science\\
  Macalester College\\
   \And
  Mariya Bessonov\\
  Department of Mathematics\\
  NYC College of Technology\\
  \And
   Korana Burke \\
   Department of Mathematics\\
   University of California Davis
  \And
  Simone Cassani \\
  Department of Mathematics\\
  University at Buffalo
 \AND
  Maria-Veronica Ciocanel \\
  Department of Mathematics and Department of Biology \\
  Duke University \\
  \And
  Daniel B. Cooney \\
  Department of Mathematics and Center for Mathematical Biology\\
  University of Pennsylvania
  \And
  Alexandria Volkening \\
  Department of Mathematics\\
  Purdue University
}

\renewcommand{\shorttitle}{How do classroom-turnover times depend on lecture-hall size?}
\renewcommand{\headeright}{}
\renewcommand{\undertitle}{}

\begin{document}
\maketitle

\begin{abstract}
Academic spaces in colleges and universities span classrooms for $10$ students to lecture halls that hold over $600$ people. During the break between consecutive classes, students from the first class must leave and the new class must find their desks, regardless of whether the room holds $10$ or $600$ people. Here we address the question of how the size of large lecture halls affects classroom-turnover times, focusing on non-emergency settings. By adapting the established social-force model, we treat students as individuals who interact and move through classrooms to reach their destinations. We find that social interactions and the separation time between consecutive classes strongly influence how long it takes entering students to reach their desks, and that these effects are more pronounced in larger lecture halls. While the median time that individual students must travel increases with decreased separation time, we find that shorter separation times lead to shorter classroom-turnover times overall. This suggests that the effects of scheduling gaps and lecture-hall size on classroom dynamics depends on the perspective---individual student or whole class---that one chooses to take.
\end{abstract}


\section{Introduction}\label{sec:intro}

Pedestrian crowds display complex, nonlinear dynamics \cite{helbing2005self}, and understanding their collective behavior is an active research area \cite{Bellomo2016,Schadschneider2018,Duives2013,Zhan2008,cristiani2014multiscale} with early empirical studies dating to over six decades ago \cite{Hankin1958}. As pedestrians move toward their destinations, they interact with each other, respond to external signals, and account for physical obstacles in their environment, such as a classroom, train, or theater. Understanding how crowds travel through these spaces has applications to design \cite{Zhan2008,helbing2005self,bartlett1973frank}, disease tracing or guidelines \cite{Romero2020,sajjadi2020social}, evacuation \cite{Xu2021,Helbing2000,lee2020speed}, crowd management \cite{porter2018pedestrian}, scheduling and route choice \cite{Hoogendoorn2004}, and other areas of societal interest. The study of pedestrian crowds is thus interdisciplinary, drawing on fields including biology, sociology, psychology, physics, engineering, computer science, statistics, and mathematics \cite{Bellomo2016,Bode2019,Helbing2009,Sieben2017}. Here we take a mathematical modeling perspective and, motivated by the widespread presence of high-enrollment classes in universities, we focus on student movement and classroom-turnover dynamics in large lecture halls. 

Empirical methods for studying pedestrian dynamics include questionnaires \cite{Sieben2017}, observations in the field \cite{Davidich2013,nilsson2009social}, and experiments in lab settings \cite{Sieben2017,Moussaid2012,Hoogendoorn2005,Seyfried2009,lee2020speed}; see \cite{Feng2021} for a recent review. Complementing these methods, mathematical approaches to crowd behavior have spanned many techniques, and we highlight the reviews \cite{Schadschneider2018,Helbing2001,Duives2013,Zhan2008,Bellomo2016,Bode2019} (and references therein) for a more extended discussion than we include here. Researchers have developed macroscopic models \cite{Carrillo2016,Hughes2002,Hughes2003,colombo2012class,bertozzi2015contagion,Burger2020} for crowd density and mesoscopic, kinetic-theory models \cite{bellomo2013microscale,kim2020coupling,festa2018kinetic,Degond2013,Henderson1971}. On the microscopic side, prior work includes cellular-automaton \cite{Burstedde2001,Varas2007,Hu2018,Kirchner2002,Blue2001,Gao2020,Davidich2013,Kirchner2003} and lattice-gas models \cite{helbing2003lattice,tajima2001scaling,Kuang2008} that treat pedestrians as individuals moving in discrete space. Off-lattice microscopic models feature differential equations for pedestrian movement and can be velocity- or acceleration-based \cite{Schadschneider2018}. Models built on the social-force concept \cite{helbing1995social,helbing2005self,Helbing2009}---the framework that we use here---are a prominent off-lattice approach that has been widely studied and adapted (e.g., \cite{zanlungo2011social,porter2018pedestrian,li2015parameter,Helbing2000,seer2014validating,johansson2007specification,Ko2013}). Researchers have also developed hybrid models \cite{Kneidl2013} and detailed agent-based approaches \cite{delcea2019increasing,liu2016agent}, as well as optimal-control and game-theoretic perspectives \cite{lachapelle2011mean,Hoogendoorn2004,Dogbe2010,Zheng2011,burger2013mean,cartee2018anisotropic,achdou2019mean,Bailo2018}. 

Mathematical modeling of crowds often centers on evacuation in venues including airplanes \cite{Kirchner2003}, theaters \cite{Gao2020,nilsson2009social}, classrooms \cite{helbing2003lattice,liu2016agent,fu2012simulation,Varas2007,zhang2008experiment,Hu2018,delcea2019increasing}, and general rooms without internal barriers \cite{Kirchner2002,takimoto2003spatio,tajima2001scaling,Burstedde2001}. By combining experiments with a lattice-gas model, Helbing \emph{et al.} \cite{helbing2003lattice} found a broad distribution of escape times for students in a small, $30$-seat classroom. Focusing on a larger room with $168$ seats, Fu \emph{et al.} \cite{fu2012simulation} simulated student movement using a cellular-automaton model. Design questions related to barrier placement, door number, aisle organization, and door width also come up naturally in evacuation studies \cite{helbing2005self}. For example, Varas \emph{et al.} \cite{Varas2007} tested how different doors affect egress in classrooms of $50$--$70$ seats, and Gao \emph{et al.} \cite{Gao2020} implemented alternative aisle configurations in a theater with $900$ seats. 

Across evacuation models and experiments, motion is unidirectional rather than bidirectional: everyone (with the possible exception of emergency workers) seeks to exit the space. The ``faster is slower" dynamic \cite{Helbing2000,Garcimartin2014} can emerge in these settings. This is a phenomenon in which evacuation slows down when pedestrians attempt to move too fast. Another common feature of unidirectional flows, the ``zipper effect" occurs when pedestrians stagger themselves as they encounter bottlenecks \cite{Hoogendoorn2005,Seyfried2009}. On the other hand, bidirectional motion arises in day-to-day environments including transit platforms \cite{Hankin1958,porter2018pedestrian}, where models can be used to help address questions surrounding efficiency, route choice, and design. In bidirectional motion, collective dynamics such as lane formation or oscillating flows have been observed empirically \cite{Sieben2017} and reproduced in models (e.g., \cite{helbing1995social,Helbing2009,
Helbing2001,moussaid2011simple,Zhang2012,Burstedde2001,Burger2020,Bailo2018}).

Lecture halls in colleges are a venue that combine questions related to evacuation, efficiency, design, and scheduling. Universities also offer unique pedestrian dynamics to investigate since classes can vary in size from under $10$ to over $500$ people. Despite the associated breadth in room size, there is generally a common scheduling gap between consecutive classes at the same institution. During this break, students from one class must exit and the next class must find their desks. This raises questions about how the time that it takes to empty and fill an academic space is related to lecture-hall size, particularly as universities continue to build or renovate large classrooms \cite{DavisNew}. With this motivation, here we address classroom-turnover dynamics in lecture halls of $200$ to $600$ desks. Unlike prior studies that have considered emergency evacuation or investigated design in rooms with a fixed capacity, we use a social-force modeling approach to elucidate how lecture halls of different sizes---and similar design---affect student movement.

\section{Model and Methods}

To better understand how lecture-hall size is related to classroom-turnover time, we develop an off-lattice model of student dynamics using the social-force approach \cite{helbing1995social,helbing2005self,Helbing2009}. In the social-force framework \cite{helbing1995social,helbing2005self,Helbing2009,Helbing2000}, pedestrian ``particles" experience acceleration toward their desired velocity, repulsive forces to account for physical obstacles in the domain or interactions with other pedestrians, and possible attractive forces. Importantly, social-force models have been able to replicate many realistic phenomena such as lane and stripe formation  \cite{helbing1995social,helbing2005self,Helbing2009} or oscillating flows at bottlenecks \cite{helbing1995social,Helbing2009}.

\begin{figure*}
\includegraphics[width=\textwidth]{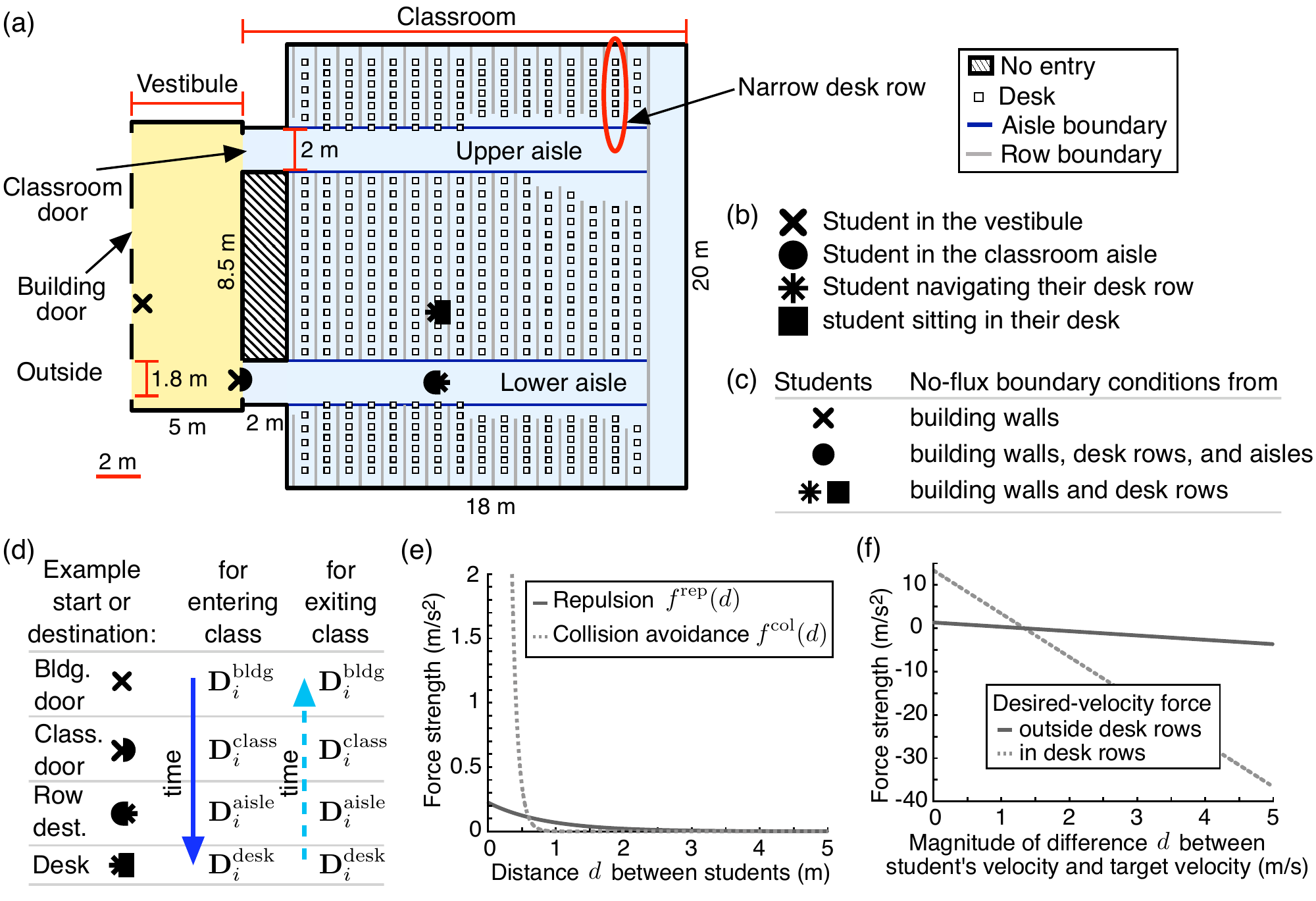}
\caption{\label{fig:modelSchematic}Model overview. (a) We model our baseline domain roughly after our measurements of Rock Hall at the University of California, Davis \cite{RockHall}. We refer to this baseline domain as ``our Rock Hall". (b) In our simulations, the symbol shape marking each student's position indicates the region of the building in which the student is located. We consider student $i$ to be in their aisle (desk row) if they are in the classroom with row status $R_i(t) = 0$ ($R_i(t)=1$); see Section~\ref{sec:modelRulesDestination}. (c) All students experience no-flux boundary conditions along the building walls and desk rows; students with a row status of zero also experience no-flux conditions along the aisles. (d) As an entering (exiting) student moves toward a desk (building door), their current destination changes. (e) Each student's movement is governed by repulsion from other students and strong local repulsion to model agents not occupying the same space. The maximum of $f^\text{col}$ is about $140$ m/s$^2$ at $d =0$. (f) Students feel a force directing them toward their destination and desired speed. To allow pedestrians to make sharp turns when they enter or exit their row of desks, we use a stronger desired-velocity force for students with $R_i(t) =1$. We plot $||\textbf{f}^\text{des}(d)||$ for $v_\text{avg} = 1.34$~m/s in Equation~\eqref{eq:desForce} \cite{waldau2007pedestrian}.}
\end{figure*}

Treating each student as an independent agent, our model accounts for entering and exiting classes and includes rules for how individuals transition between different destinations, such as a door or a desk. In Sections \ref{sec:modelSpace} and \ref{sec:modelDynamics}, we describe how we model lecture-hall spaces and student behavior, respectively. Figure~\ref{fig:modelSchematic} and Table~\ref{tab:notation} summarize our model and notation. Our code will be made available on GitLab \cite{code} and our model parameters are given in Table~\ref{tab:parameters} and in the Supplementary Material.

\begin{figure*}
\includegraphics[width=\textwidth]{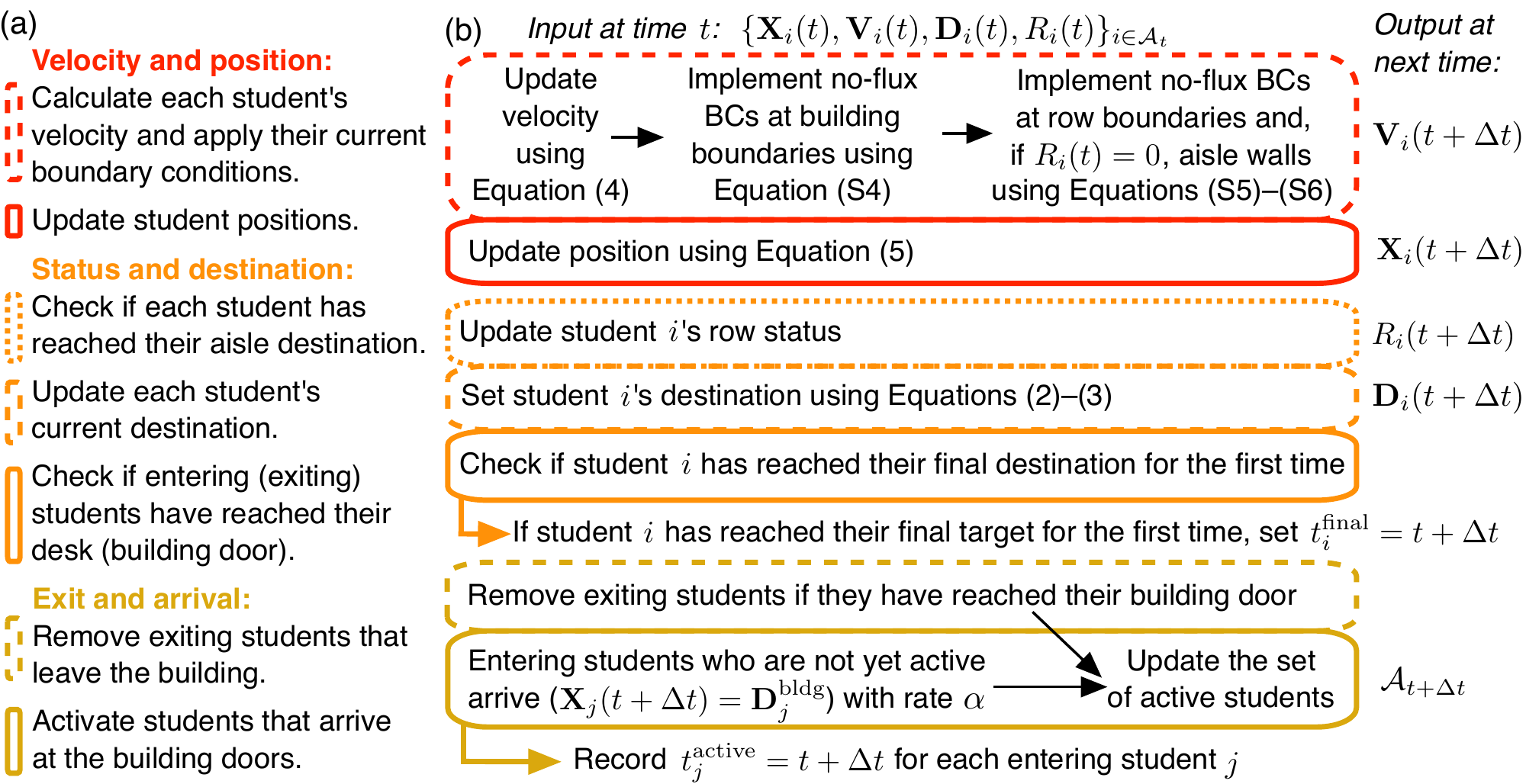}
\caption{Simulation flowchart. (a) To implement one time step $\Delta t$, we follow three main steps. (b) First, we synchronously update the velocity and then the position of each active student (i.e., student in the building) using Equations \eqref{eq:modelV}--\eqref{eq:modelX}. Second, we check if each student $i$'s row status has changed; update each student's current destination using Equations~\eqref{eq:destination}--\eqref{eq:destination2}; and log the time $t^\text{final}_i$ if student $i$ has just reached their final destination. Third, we determine if any new students have arrived to the building; if so, each of these students $j$ is activated (i.e., joins $\mathcal{A}_{t+\Delta t}$) and $t^\text{active}_j$ is logged. Similarly, we check if any exiting students have reached a building door; if so, we inactivate these students. Equations~(S5)--(S6) are in Supplementary Material.}
\label{fig:modelFlow}
\end{figure*}

\subsection{Lecture-Hall Domains}\label{sec:modelSpace}

Large universities in the United States frequently have classes with $300$ or more students (e.g., see \cite{OhioStateReport} and Supplementary Figure~1), and institutions have continued to build or renovate large lecture halls \cite{DavisNew}. We construct our baseline domain after Rock Hall at the University of California, Davis. Rock Hall has functioned as a movie theater and a concert hall, but it is now used mainly for high-enrollment classes. 
Rock Hall (Supplementary Figures~1(c)--(d)) can accommodate $416$ people and is a good example of a generic large classroom. As we show in Figure~\ref{fig:modelSchematic}(a), our baseline domain---i.e., ``our Rock Hall"---has a vestibule with four external doors and a classroom with two doors and two aisles. 

In Section \ref{sec:size}, we investigate how lecture-hall size affects classroom-turnover times. To do so, 
we adapt our baseline domain to construct smaller and larger rooms, spanning $200$ to $600$ desks (Supplementary Figure~2). Because our focus is on size rather than design, we make minimal changes to the room geometry and instead accommodate more students by adding more rows of desks. See Supplementary Material for more details.

\subsection{Student Dynamics}\label{sec:modelDynamics}

We model each student as an agent with seven intrinsic features and four time-dependent variables as below:
\begin{align}
    \text{student $i$} =& \left\{\underbrace{C_i,\tau^\text{pre}_i, v^\text{des}_i, \textbf{D}^\text{bldg}_{i}, \textbf{D}^\text{class}_{i}, \textbf{D}^\text{aisle}_{i}, \textbf{D}^\text{desk}_{i}}_\text{time-independent features} \right. \, ,\nonumber \\ &
  \hspace{7mm}  \left.\underbrace{\textbf{X}_i(t), \textbf{V}_i(t), \textbf{D}_i(t), R_i(t)}_\text{time-dependent variables}\right\},
\end{align}
where $C_i$ denotes student $i$'s class (entering or exiting); $\tau^\text{pre}_i$ is the time that it takes student $i$ to prepare to leave their desk (if they are in the exiting class); $v^\text{des}_i$ is their desired speed; $\textbf{D}^\text{bldg}_{i}, \textbf{D}^\text{class}_{i}$, $\textbf{D}^\text{aisle}_{i}$, and $\textbf{D}^\text{desk}_i$ correspond to student $i$'s desired building door, classroom door, aisle destination, and desk, respectively; 
$\textbf{X}_i(t)$ is their position; $\textbf{V}_i(t)$ is their velocity; $\textbf{D}_i(t)$ is their current destination; and $R_i(t)$ is a status variable that logs whether or not student $i$ is in a desk row. Figure~\ref{fig:modelFlow} overviews how we update student $i$'s time-dependent features.

We assume that there are $N^\text{enter} \ge 0$ and $N^\text{exit} \ge 0$ students in the entering and exiting classes, respectively. We let $\mathcal{A}_t \subset \{1,..., N^\text{enter} + N^\text{exit}\}$ be the set of students who are in the building (or ``active") at time $t$, and we only track this set of active students in our simulations (Figure~\ref{fig:modelSchematic}(a)). Our model of student behavior then includes three main components: differential equations for agent movement in lecture-hall domains (Section \ref{sec:modelMovement}), discrete-time rules for arrival and exit at the building doors (Section \ref{sec:modelArrivalExit}), and discrete-time rules that govern how each student $i$'s current target $\textbf{D}_i(t)$ is selected from among $\textbf{D}^\text{bldg}_i$, $\textbf{D}^\text{class}_i$, $\textbf{D}^\text{aisle}_i$, and $\textbf{D}^\text{desk}_i$ and how their row status $R_i(t)$ evolves (Section \ref{sec:modelRulesDestination}). As initial conditions, we assume that each student $i$ in the exiting class starts at a desk with zero velocity. Once active, entering students begin with zero velocity at one of the building doors. We also allow $N^\text{early} \ge 0$ entering students to start in the vestibule. See Supplementary Material for details about our initial conditions.

To account for some variability, we assign each student~$i$ a pre-movement time $\tau^\text{pre}_i$ and desired speed $v^\text{des}_i$. We draw the desired speeds from a normal distribution with mean $v_\mathrm{avg} = 1.34$ meters per second (m/s) and standard deviation $v_\mathrm{std} = 0.37$ m/s; these values are based on prior studies \cite{helbing1995social,helbing2005self,Ko2013,Moussaid2012,Zhang2012,waldau2007pedestrian}. To our knowledge, data on the time that it takes people to gather their belongings and prepare to move after a lecture in non-emergency settings is limited. Notably, Zhang \emph{et al.} \cite{zhang2008experiment} track pre-movement times in classrooms during evacuation experiments. The study \cite{nilsson2009social} also provides pre-movement times for evacuations in a theater. Motivated in part by the distributions in \cite{nilsson2009social}, we sample $\tau^{\text{pre}}_i$ for exiting students from a normal distribution with mean $\mu_\text{pre} = 35$~s and standard deviation $\sigma_\text{pre} = 20$~s. We set $\tau^\text{pre}_i = 0$ for entering students. In practice, we truncate the distribution for $v^\text{des}_i$ at one standard deviation from the mean, and we implement cutoffs for $\tau^\text{pre}_i$ to ensure this delay time is positive and no longer than $2$ minutes.

\begin{table*}[t]
\setlength\tabcolsep{2pt}
   \center
        \caption{Summary of our notation; see Figures \ref{fig:modelSchematic} and \ref{fig:modelFlow} for a model overview and flowchart, respectively.}
    \begin{tabular}{l p{15.9cm}}
    \textbf{Notation} &  {\textbf{Meaning}}  \\ 
    \hline 
    $\mathcal{A}_t$ & set of students that are in the building (i.e., active) at time $t$  \\
    $\tau^\text{pre}_i$ & time that exiting student $i$ takes to prepare to leave their desk (if $i$ is in the entering class, $\tau^\text{pre}_i =0$) \\
    $v^\text{des}_i$ & student $i$'s desired speed \\
    $\textbf{D}^\text{bldg}_{i}$ & entering student $i$'s initial condition or exiting student $i$'s final target (a building door to outside; see Figure~\ref{fig:modelSchematic}(d))\\
   $\textbf{D}^\text{class}_{i}$ & coordinates of internal door (between vestibule and classroom) that student $i$ uses\\
    $\textbf{D}^\text{aisle}_{i}$ & student $i$'s desired destination in the aisle near their row of desks  \\
    $\textbf{D}^\text{desk}_{i}$ & exiting student $i$'s initial condition or entering student $i$'s final target (a desk in the classroom) \\
    $\textbf{X}_i(t)$ & position of student $i$ at time $t$ \\
        $\textbf{V}_i(t)$ & velocity of student $i$ at time $t$ \\
    $\textbf{D}_i(t)$ & current destination of student $i$ at time $t$ ($\textbf{D}^\text{bldg}_i$, $\textbf{D}^\text{class}_{i}$, $\textbf{D}^\text{aisle}_{i}$, or $\textbf{D}^\text{desk}_{i}$, depending on their location and class) \\
    $R_i(t)$ & row status of student $i$ at time $t$ ($R_i(t) = 1$ if student $i$ is in their narrow desk row and $0$ otherwise) \\
    $t^\text{active}_i$ & time that student $i$ is first active (in the building) \\
    $t^\text{final}_i$ & time that student $i$ first reaches their final destination \\
    \hline
    \end{tabular}
    \label{tab:notation}
\end{table*}

\subsubsection{Student Destinations} \label{sec:modelRulesDestination}

In our model, entering students become active at building doors and have a final goal of reaching the position of their desk ($\textbf{D}^\text{desk}_i$). Each exiting student $i$ starts at a desk and has a building door ($\textbf{D}^\text{bldg}_i$) from which they seek to exit. At the start of a simulation, we assign desks and building doors to pedestrians uniformly at random from the set of desks and four building doors, respectively; see Figure~\ref{fig:modelSchematic}(a). However, if a student immediately moved toward their desk when they entered the building, they would attempt to travel through walls. We thus introduce two intermediate destinations, and student $i$'s current target $\textbf{D}_i(t)$ changes as they move through space.

As we show in Figure~\ref{fig:modelSchematic}(d), if student $i$ is in the entering class and has arrived to the building, their first destination is $\textbf{D}^\text{class}_{i}$, corresponding to the internal door that they will use to enter the classroom. Their next target is $\textbf{D}^\text{aisle}_i$, the location in the center of the aisle associated with their desired row of desks. Once student $i$ has reached this ``aisle destination", their final target is the coordinates of their desk, namely $\textbf{D}^\text{desk}_{i}$. Similarly, if student $i$ is in the exiting class, their first target is $\textbf{D}^\text{aisle}_i$. Once they have exited their row of desks, student~$i$ seeks to move toward their classroom door ($\textbf{D}^\text{class}_i$) and then their building door ($\textbf{D}^\text{bldg}_i$). Importantly, $\textbf{D}^\text{class}_i$ and $\textbf{D}^\text{aisle}_i$ depend on $\textbf{D}^\text{desk}_i$: we assume students use the classroom door and travel in the aisle closest to their desk. See Supplementary Material for details.

To select $\textbf{D}_i(t)$ from $\textbf{D}^\text{bldg}_i$, $\textbf{D}^\text{class}_i$, $\textbf{D}^\text{aisle}_i$, and $\textbf{D}^\text{desk}_i$, we introduce a status variable that tracks whether or not student $i$ has reached $\textbf{D}^\text{aisle}_i$, as below:
\begin{align*}
R_i(t) &= \begin{cases} 1 & \text{if student $i$ is in their narrow desk row} \\
0 & \text{otherwise}.
\end{cases}
\end{align*}
The row status of each student in the entering (exiting) class starts at zero (one) and changes to one (zero) once they reach their aisle destination. Student $i$ has reached their aisle destination when $||\textbf{X}_i(t) - \textbf{D}_i^\text{aisle}|| < d_\text{tol}$, where $d_\text{tol} = 0.3$~m is a tolerance distance and $|| \textbf{x}|| = \sqrt{x_1^2 + x_2^2}$. We then update $\textbf{D}_i(t)$ as follows:
\begin{align}
    \textbf{D}_i(t) &= \begin{cases}
    \textbf{D}^\text{class}_{i} & \text{if $i$ is not inside the classroom } \\
    \textbf{D}^\text{aisle}_{i} & \text{if $i$ is in the classroom and $R_i(t) =0$} \\
    \textbf{D}^\text{desk}_{i} & \text{otherwise.} \end{cases}\label{eq:destination}
\end{align}
See Figure~\ref{fig:modelSchematic}(a). Similarly, for each student $i$ in the exiting class, we specify their current destination as below:
\begin{align}
    \textbf{D}_i(t) &= \begin{cases}
    \textbf{D}^\text{aisle}_{i} & \text{if $R_i(t) = 1$ ($i$ is in their desk row)} \\
    \textbf{D}^\text{class}_{i} & \text{if $R_i(t) = 0$ and $i$ is in the classroom} \\
    \textbf{D}^\text{bldg}_{i} & \text{otherwise.}
    \end{cases}\label{eq:destination2}
\end{align}
We update $R_i(t)$ and $\textbf{D}_i(t)$ as appropriate at each time step $\Delta t = 0.01$~s. As we discuss next in Section~\ref{sec:modelMovement}, $\textbf{D}_i(t)$ and $R_i(t)$ play a role in student movement.

\subsubsection{Student Movement}\label{sec:modelMovement}

We describe student movement through a social-force modeling approach \cite{helbing1995social,helbing2005self,Helbing2009}. This model treats pedestrians as particles that react to social forces. Pedestrians do not want to bump into other agents, so they feel a repulsive force from other individuals. At the same time, pedestrians experience attractive forces toward their destinations. For each active student $i \in \mathcal{A}_t$, the evolution of their position $\textbf{X}_i(t)$ and velocity $\textbf{V}_i(t)$ is given by:

\begin{align} 
d\textbf{V}_i(t) &= \begin{cases} \underbrace{\sum_{j \in \mathcal{A}_t,j\neq i} \textbf{f}^\text{col}(\textbf{x}_{ij})dt}_{\text{collision avoidance}} + \underbrace{\sum_{j \in \mathcal{A}_t,j\neq i} \textbf{f}^\text{rep}(\textbf{x}_{ij}, \textbf{v}_{ij})dt}_{\text{student repulsion}}
+ \underbrace{\bold{f}^\text{des}(\textbf{D}_i - \textbf{X}_i, \textbf{V}_i, v^\text{des}_i, R_i)dt}_\text{desired velocity} + \underbrace{\sigma d\textbf{W}^i(t)}_\text{stochasticity} & \text{if $t \ge \tau^\text{pre}_i$} \\
0 & \text{otherwise}, \label{eq:modelV} 
\end{cases}
\end{align}
\begin{align}
\frac{d\textbf{X}_i}{dt}(t) &= \textbf{V}_i(t), \label{eq:modelX}
\end{align}

where $\textbf{W}^i$ is a Wiener process; $\sigma$ is the noise strength; $\tau^\text{pre}_i$ is student $i$'s pre-movement time ($\tau^\text{pre}_i = 0$ if $i$ is in the entering class); and the forces $\bold{f}^\text{col}$ and $\bold{f}^\text{rep}$ depend on the difference between student positions (centers of mass) $\textbf{x}_{ij} = \textbf{X}_i - \textbf{X}_j$ and between their velocities $\textbf{v}_{ij} = \textbf{V}_i - \textbf{V}_j$.

The force $\bold{f}^\text{col}$ operates strongly over a short distance to prevent collisions, and $\bold{f}^\text{rep}$ acts weakly over a longer range to help students maintain a comfortable distance from one another. These forces can be expressed in terms of potentials $U^{\text{rep}}$ and $U^\text{col}$ (see Equation~\eqref{eq:baseForce}) as below:
\begin{align*}
    \bold{f}^\text{col}(\textbf{x}) & = -\nabla_{\textbf{x}}U^\text{col}(\xi_\text{C}(\textbf{x})), \\
    \bold{f}^\text{rep}(\textbf{x}, \textbf{v}) &= -\nabla_\textbf{x} U^\text{rep}(\xi_\text{E}(\textbf{x},\textbf{v}\delta_t)),
\end{align*}
where $\xi_\text{C}(\textbf{x}_{ij})$ and $\xi_\text{E}(\textbf{x}_{ij},\textbf{v}_{ij} \delta_t)$ correspond to the perceived distance between individuals $i$ and $j$. These distances can depend on the current difference between their positions ($\textbf{x}_{ij}$) and on the projected difference in their future positions ($\textbf{x}_{ij} + \textbf{v}_{ij} \delta_t$) over a timescale $\delta_t$, assuming that $\textbf{v}_{ij}$ is constant on the time interval $[t,t + \delta_t)$. 

Following the example of \cite{Helbing2009, johansson2007specification}, we use two methods for specifying the interaction forces between students: one depends only on the current distance between their positions (``the circular force specification") and the other accounts for their current and projected future positions (``the elliptical force specification"). The circular distance is $\xi_C(\textbf{x}) = || \textbf{x}||$, and the elliptical distance is given by:
\begin{align} 
   \xi_\text{E}(\textbf{x}, \textbf{v} \delta_t) &= \frac{\sqrt{\left(||\textbf{x}|| + || \textbf{x} + \textbf{v} \delta_t|| \right)^2 - ||\textbf{v}\delta_t||^2}}{2} \label{eq:ellipticaldistance}.
\end{align}

We define the repulsive forces in our model by taking the gradient of the potentials $U^\text{***}$ with respect to $\textbf{x}$ \cite{helbing1995social}, which yields the following expressions:
\begin{align*} 
    \bold{f}^{\text{col}}(\textbf{x}) &= -\nabla_{\textbf{x}} U^{\text{col}}(\xi_\text{C}(\textbf{x})) = f^\text{col}(\xi_\text{C}(\textbf{x})) \nabla_\textbf{x} \xi_\text{C}(\textbf{x}),\\
    \bold{f}^{\text{rep}}(\textbf{x},\textbf{v}) &= -\nabla_{\textbf{x}} U^{\text{rep}}(\xi_\text{E}(\textbf{x},\textbf{v}\delta_t)) \\
    &= f^\text{rep}(\xi_\text{E}(\textbf{x},\textbf{v}\delta_t)) \nabla_\textbf{x} \xi_\text{E}(\textbf{x},\textbf{v}\delta_t),
\end{align*}
where 
the gradients of $\xi_{*}(\cdot)$ are given by:

\begin{align}
   \nabla_{\textbf{x}} \xi_\text{C}(\textbf{x}) &= \frac{\textbf{x}}{||\textbf{x}||} \label{eq:circGrad}\\ 
    \nabla_{\textbf{x}} \xi_\text{E}(\textbf{x}, \textbf{v} \delta_t) &= \frac{|| \textbf{x}|| + || \textbf{x} +  \textbf{v} \delta_t || }{2 \sqrt{\left(||\textbf{x}|| + ||\textbf{x} +  \textbf{v} \delta_t||\right)^2 - || \textbf{v} \delta_t||^2}}  \left( \frac{\textbf{x}}{|| \textbf{x}||} + \frac{\textbf{x}+   \textbf{v} \delta t}{|| \textbf{x}+ \textbf{v} \delta_t||}\right),\label{eq:ellipGrad}
\end{align}

and
\begin{align}
    f^{***}(x) &= B^{***}e^{(r-x)/b^{***}}.\label{eq:baseForce}
\end{align}
Here $B^{***}$ is related to the force strength; $b^{***}$ is the interaction range; $^{***} \in \{\text{col}, \text{rep} \}$ indicates that we use different parameters \cite{johansson2007specification} for each force (see Figure~\ref{fig:modelSchematic}(e) and Table~\ref{tab:parameters}); and $r =0.6$~m is the diameter of the privacy sphere that pedestrians seek to maintain \cite{helbing2005self}. When $\delta_t  = 0$ s, Equation~\eqref{eq:ellipGrad} simplifies to Equation~\eqref{eq:circGrad} and $\bold{f}^\text{rep}$ has the same form as $\bold{f}^\text{col}$. The elliptical model specification ($\delta_t > 0$)  allows pedestrian interactions to depend on relative velocity as well as on distance, and it enables the repulsive force to have a lateral component. 
As a consequence, this approach leads to smoother, less confrontational pedestrian behavior \cite{johansson2007specification,Helbing2009}. 
We use $\delta_t = 0.1$~s, which is similar to values used in \cite{Helbing2009}. 

Lastly, to account for each student's current destination, we specify that agent $i \in \mathcal{A}_t$ feels an attractive force \cite{helbing1995social,johansson2007specification} in the direction of $\textbf{d} = \textbf{D}_i(t) -\textbf{X}_i(t)$:
\begin{align}
    \bold{f}^\text{des}(\textbf{d},\textbf{v},v^\text{des},R) &= \begin{cases}
    \displaystyle\frac{1}{\tau}\left(v^\text{des} \frac{\textbf{d}}{||\textbf{d}||}- \textbf{v}\right) & \text{if $R \neq 0$ }\\
   \displaystyle\frac{1}{\tau_\text{row}}\left(v^\text{des}\frac{\textbf{d}}{||\textbf{d}||} - \textbf{v}\right) & \text{if $R = 0$ }
    \end{cases}, \label{eq:desForce}
\end{align}
where $\{\tau, \tau_\text{row}\}$ are the relaxation times in which students adapt their velocity $\textbf{V}_i(t)$ to move toward their current destination $\textbf{D}_i(t)$ with speed $v$; $R\neq 0$ means the student is not in a desk row; and $ R = 0$ means the student is in a desk row. Similar to prior studies \cite{helbing1995social,helbing2005self}, we choose a baseline relaxation time of $\tau = 1$~s. However, we find that students need to make sharper adjustments to their velocity when turning into (or out of) their desk row and navigating the narrow rows of the classroom. Following the example in \cite{helbing1995social}, we thus set $\tau_\text{row} = 0.1$~s for entering (exiting) students who have reached (not yet left) the row of their desired desk.

We implement no-flux boundary conditions along the building outline. We also treat the boundaries of the desk rows as physical walls to model desks that are too close together for students to move between rows; see Figure~\ref{fig:modelSchematic}. Additionally, if $R_i(t) = 0$, we include no-flux boundary conditions along the aisle boundaries. This models our assumption that entering students do not want to step out of the aisle unless they have arrived at their desk row. See Supplementary Material for more information.

\subsubsection{Student Arrival and Exit}\label{sec:modelArrivalExit}

At each time step $\Delta t$, if the entire entering class is not yet in the building, students arrive at the doors stochastically with rate $\alpha$. In particular, if entering student $i \notin \mathcal{A}_t$, then $i \in \mathcal{A}_{t +\Delta t}$ with probability
\begin{align}
p &= \frac{\alpha \Delta t}{\text{number of entering students not yet active}}.\label{eq:probActive}
\end{align}
Once all of the entering students are active, no new students enter the building. We define the time that each student $i$ first becomes active as
\begin{align}
    t^\text{active}_i &= \text{earliest time that $i$ is in the building},
\end{align}
and we note that $t^\text{active}_i =0$~s for each exiting student~$i$, since these students begin in the classroom. For our simulations with only an entering class, we also include $N^\text{early} >0$ students who start active in the vestibule; when we include an exiting class, we use $N^\text{early} = 0$. 

Students from the exiting class become inactive and are removed from our simulation when they are within a tolerance distance $d_\text{tol}$ of their assigned building door ($\textbf{D}^\text{bldg}_i$ for student $i$) or have left the building. Entering students, on the other hand, are active from the time of their arrival onward. We define 
\begin{align}
    t^\text{final}_i &= \text{time that $i$ reaches their final destination},\label{eq:finalTime}
\end{align}
where this is the time that entering students first reach their desks and exiting students exit the building (or reach a distance $d_\text{tol}$ from their desired building door). If a student does not reach their final target during our simulation, we set $t^\text{final}_i = t_\text{max} + 1$, where $t_\text{max}$ is the total simulation time.

\section{Results}

Using our model from Section~\ref{sec:modelDynamics}, we now present a study of the effects of lecture-hall size on classroom-turnover times. We discuss our results in terms of two time quantities, both measured in seconds: the simulation time $t$ and the travel time of individual students. When there is an exiting class present, $t=0$~s is the time when this first class ends. Understanding how long it takes all of the entering students to reach their desks measured from $t=0$~s thus provides information about classroom-turnover times. From a single student $i$'s perspective, we also find it useful to define their travel time from building door to desk (or vice versa), namely
\begin{align}
    t^\text{trav}_i &= t^\text{final}_i - t^\text{active}_i.\label{eq:travTime}
\end{align}

In Section~\ref{sec:baseline}, we begin with a baseline study of how quickly an entering (exiting) class of students can reach their desks (leave the building) in the absence of another class. This models dynamics in the first or last class of the day, respectively. In Section~\ref{sec:bidirectional}, we consider two classes, scheduled consecutively, in the same lecture hall. Lastly, in Section~\ref{sec:size}, we test the effects of lecture-hall size on classroom-turnover times. Our results indicate that lecture-hall size and the time between consecutive classes work together to control how long it takes students to exit the classroom or reach their desks.

\subsection{Baseline: Our Rock Hall with One Class}\label{sec:baseline}

\begin{figure*}
\includegraphics[width=\textwidth]{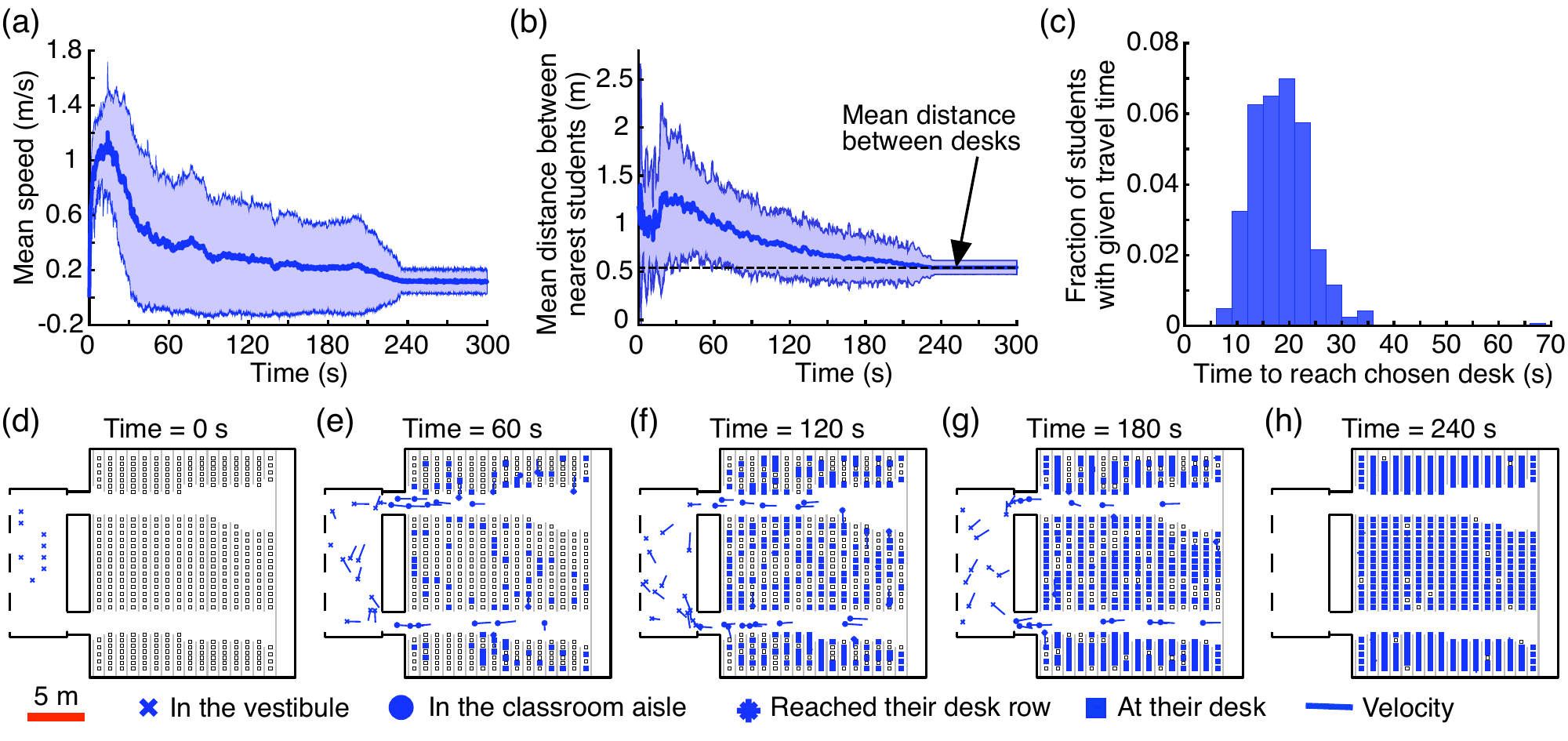}
\caption{\label{fig:entering_onesim}Example simulation of $400$ students entering our Rock Hall, with no exiting class. (a) We show the mean speed of active students through time for a representative simulation. (b) As time (from the start of the simulation) passes, the mean distance between nearest students approaches the average distance between desks. In panels (a)--(b), solid lines denote the mean, and shaded areas indicate the standard deviation over all active students at each time point. (c) The distribution for the time (i.e., $t^\text{trav}_i$ for student $i$) that it takes students to travel from a building door to their desk is right-skewed. Our histogram bin width is $3$~s. (d)--(h) We show positions and velocities of students at five time points for an example simulation.}
\end{figure*}

To establish a baseline for our model, we first simulate $400$ students entering our Rock Hall domain, with no exiting class. Figure~\ref{fig:entering_onesim} shows time snapshots of students entering the classroom, as well as summary statistics of pedestrian dynamics. The students' mean speed is close to their average desired velocity in the vestibule, and it decreases with time as students reach their desks (Figure~\ref{fig:entering_onesim}(a)). Similarly, students are relatively spread out in the vestibule, but the mean distance between nearest-neighboring students eventually decreases to the distance between desks (Figure~\ref{fig:entering_onesim}(b)). We provide the distribution of student travel times $t^\text{trav}_i$ in Figure~\ref{fig:entering_onesim}(c).

\begin{figure*}
\includegraphics[width=\textwidth]{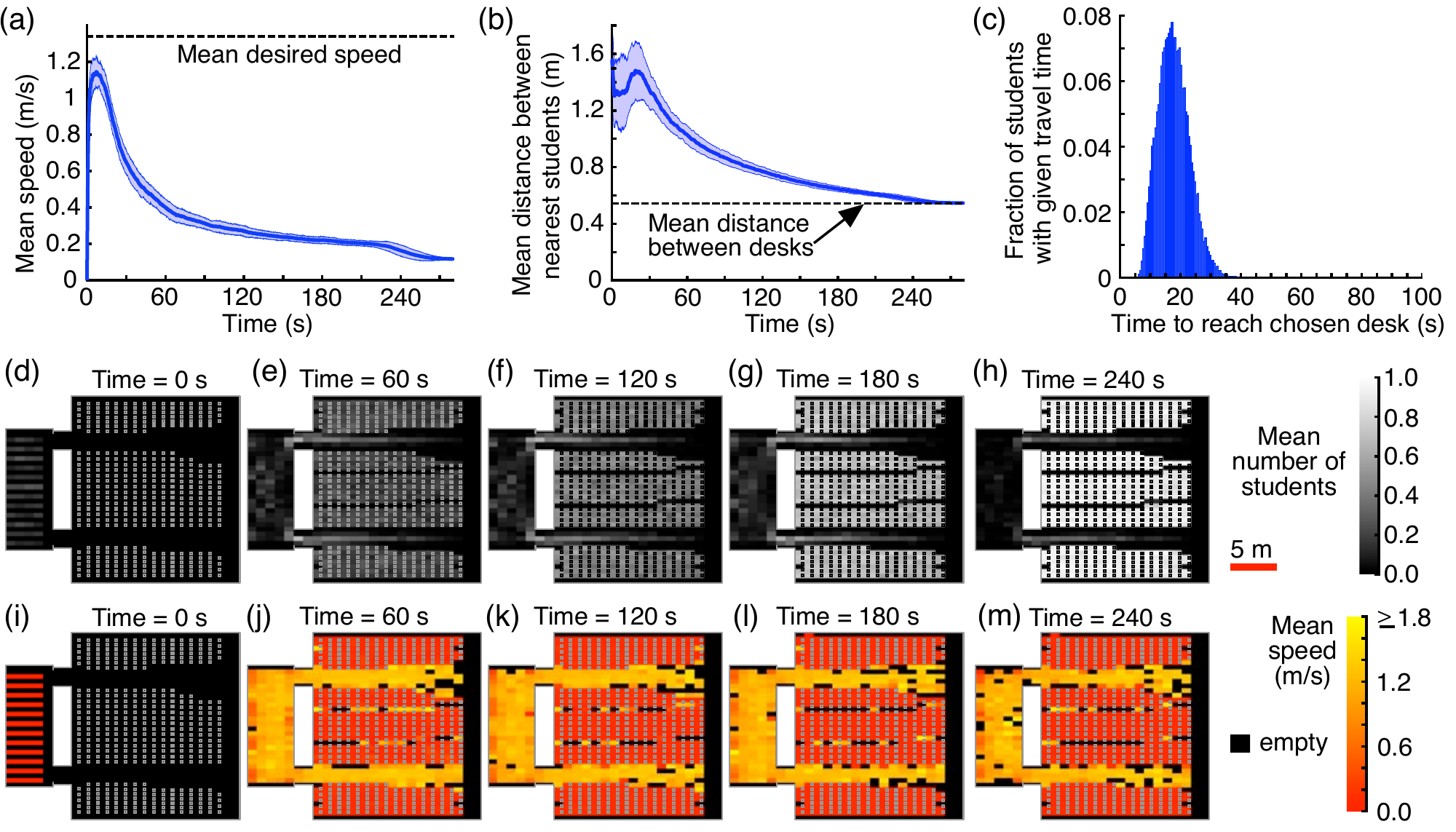}
\caption{\label{fig:entering_avgsims}Entering class of $400$ students in our Rock Hall, with no exiting class. All panels present summary statistics across $100$ simulations.
In agreement with Figure~\ref{fig:entering_onesim}, the (a) mean speed and (b) mean distance between nearest students decays in time. In panels (a)--(b), solid (shaded) lines are the mean (standard deviation) across $100$ simulations. (c) We show the travel-time distribution for all students across $100$ simulations. The histogram bin width is $0.5$ s, and we crop the long tail in the distribution at $100$~s; a few outlier agents (about $0.9$\% of the total number of students across $100$ simulations) take a longer time to reach their desks. For comparison with Figures~\ref{fig:entering_onesim}(d)--(h), we show in (d)--(h) mean mass (number of students in grid squares with area $1.5$ m$^2$) and (i)--(m) associated mean speed in those grid squares across $100$ simulations at five time points; see Supplementary Material. Each grid square contains at most one desk, and desk positions are indicated as gray or black squares. Note $N^\text{entering} >0$ students start in the vestibule, visible in panels (d) and (i). In panels (l)--(m), most students have reached their desks, so the velocities are means across a limited number of agents.} 
\end{figure*}

Since our model is stochastic, we also investigate if these trends are consistent across multiple simulations. In Appendix~\ref{app:uncertainty}, we describe the A--test method for uncertainty quantification based on \cite{vargha2000critique}. Using this method, we find that $100$ simulations of our model are sufficient to mitigate the uncertainty introduced by stochastic model elements on the mean student travel time. For the remainder of our paper, we therefore show summary statistics based on $100$ simulations for our results, with the exception of sample simulation snapshots. 

Figures~\ref{fig:entering_avgsims}(a)--(c) display speeds and student--student distances across $100$ simulations of $400$ entering students in our Rock Hall. Across our $100$ simulations, all of the students reach their desks by the total simulation time ($t_\text{max} = 280$~s). However, we find that $36$ students take longer than $100$~s to reach their desk, which can occur when they enter the wrong desk row. This amounts to $0.9\%$ of the total number of students in these simulations, and as a result we crop the long tail in our travel-time distributions in Figure~\ref{fig:entering_avgsims}(c). We similarly observe very low percentages of stuck pedestrians in other simulation settings, where we again omit these students from our figures; see Supplementary Material for details.

\begin{figure*}
\includegraphics[width=\textwidth]{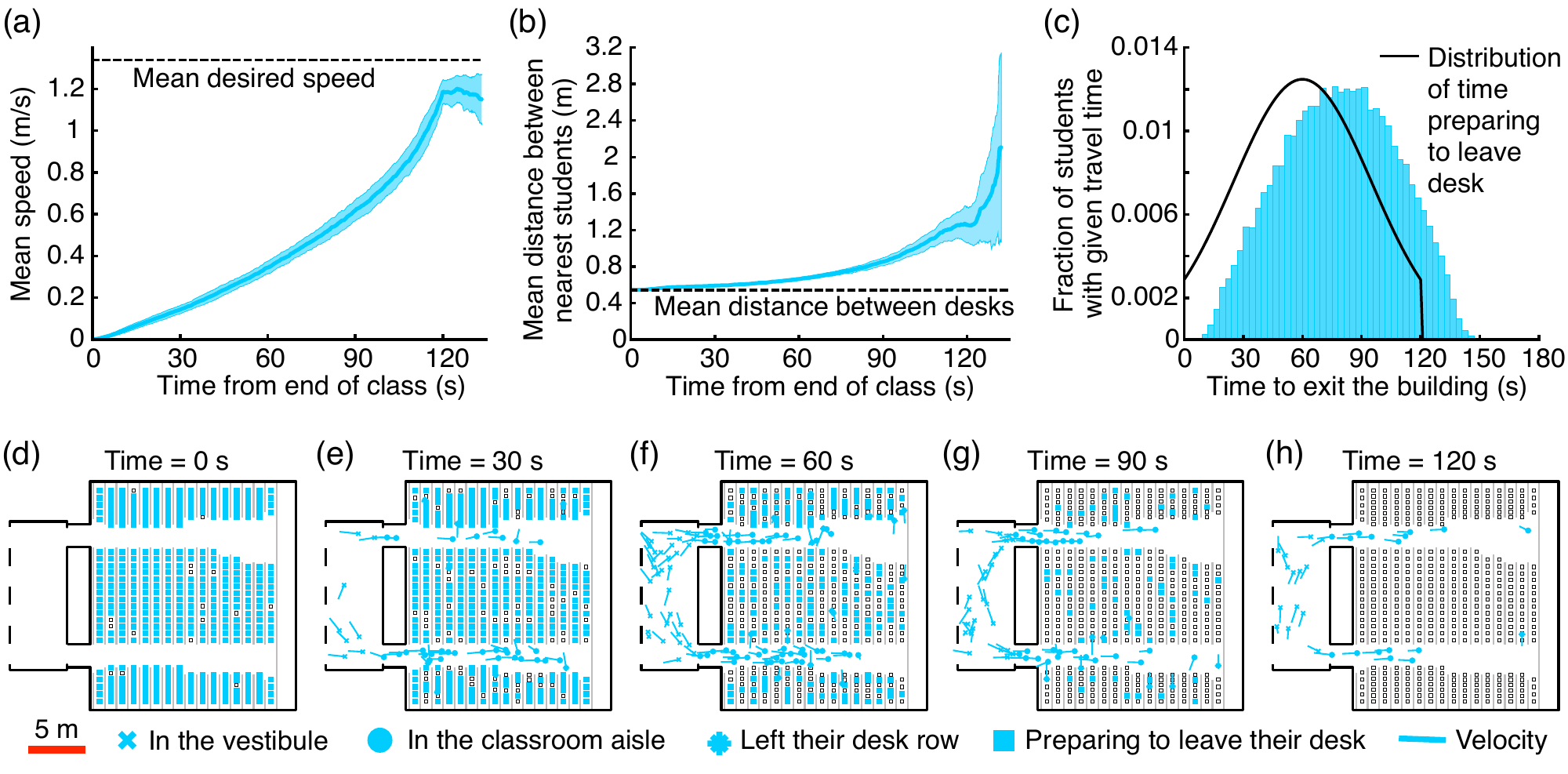}
\caption{\label{fig:exit_avgsims}Exiting class of $400$ students in our Rock Hall, with no entering class. Panels (a)--(c) present summary statistics across $100$ simulations. (a) Students start with zero velocity at their desks and approach the mean desired speed in time. (b) As time passes, more students depart the building and the standard deviation in student--student distances grows. In panels (a)--(b), solid (shaded) lines indicate the average (standard deviation) over $100$ simulations. (c) The distribution for the time that it takes students to exit the building after class ends has a similar shape as our distribution for the time ($\tau_i^\text{pre}$ for student $i$) that it takes students to gather their belongings and leave their desk. Histogram bin width is $3$~s. (d)--(h) In this example simulation, blue squares denote students who are active but still gathering their belongings; these students can exert forces on others but do not feel forces themselves until $t\ge \tau^\text{pre}_i$; see Equation~\eqref{eq:modelV}.}
\end{figure*}

We next simulate $400$ students exiting Rock Hall, with no entering class. Figure~\ref{fig:exit_avgsims} shows summary statistics across $100$ simulations, as well as time snapshots for an example simulation. The mean speed of exiting students in Figure~\ref{fig:exit_avgsims}(a) behaves opposite to that of entering students in Figure~\ref{fig:entering_avgsims}(a). Students initially move slowly as they navigate tight rows and crowded aisles, and they later approach their desired speed. The mean distance between neighboring students is initially determined by the spacing between desks, but it increases as students leave their desks (Figure~\ref{fig:exit_avgsims}(b)). As more students exit the building, the standard deviation in student--student distance increases, since it is calculated over a smaller number of agents. The time that it takes students to exit appears to roughly follow a normal distribution; see Figure~\ref{fig:exit_avgsims}(c). As we discuss in Section \ref{sec:modelDynamics}, we model the amount of time $\tau_i^{\mathrm{pre}}$ that students take to gather their belongings by a truncated normal distribution with parameters in Table~\ref{tab:parameters}. The travel-time distribution is, as we would expect, a shifted version of this distribution of pre-movement times.

\subsection{Our Rock Hall with Both Entering and Exiting Classes} \label{sec:bidirectional}

Scheduling consecutive classes in the same lecture hall is likely to impact student travel times. To investigate this, we consider bidirectional simulations in our Rock Hall. Specifically, we model $N^\mathrm{exiting}=400$ students who begin to exit the lecture hall at the end of their class, and $N^\mathrm{entering}=400$ students who start arriving to the building for the next class. We specify a separation time $t_\text{sep} = 90$~s between the end of the first class and the time when students in the next scheduled class begin to arrive at the building doors. For all of our simulations with both entering and exiting classes, we set $N^\text{early} =0$, so there are no students who start in the vestibule.

\begin{figure*}
    \centering
    \includegraphics[width=\textwidth]{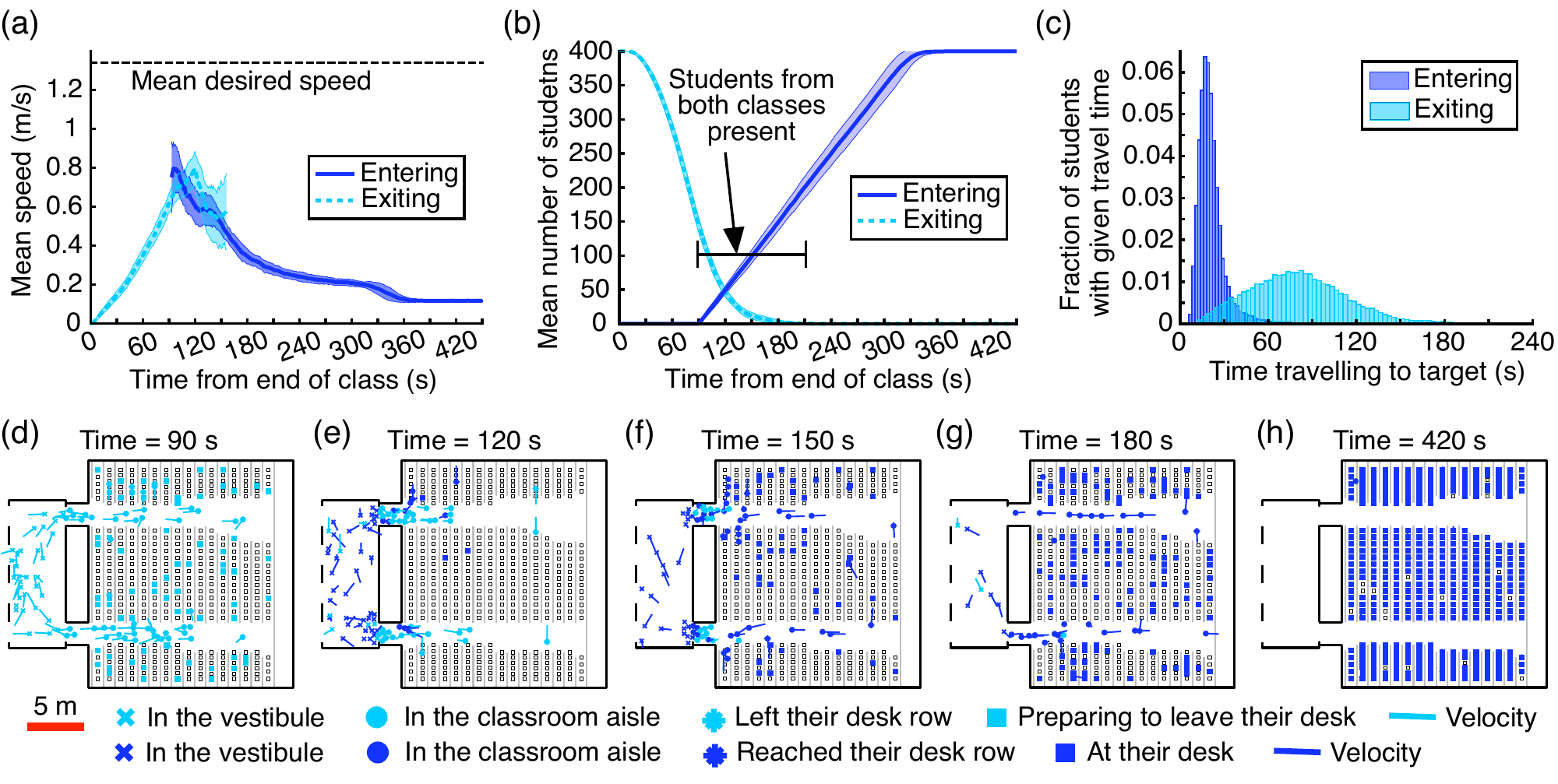}
    \caption{\label{fig:baselineBidir}Consecutive classes of $400$ students each in our Rock Hall, with one class exiting and the next class entering. Panels (a)--(b) present summary statistics across $100$ simulations. (a) We indicate the mean desired speed (dashed line; see Equation~\eqref{eq:desForce}) and the mean speed for each class of students through time. We assume entering students do not begin arriving at the building doors until $t_\text{sep} = 90$~s after the end of the first class. (b) As time passes, fewer exiting students remain, and more entering students become active. In panels (a)--(b), solid (shaded) lines are the average (standard deviation) over $100$ simulations. (c) The travel-time distributions have longer tails than in the entering-only (Figure~\ref{fig:entering_avgsims}(c)) and exiting-only (Figure \ref{fig:exit_avgsims}(c)) simulations. Histogram bin widths are $2$~s and $3$~s for entering and exiting students, respectively. (d)--(h) In this example simulation, pedestrians interact with students from the other class mainly in the vestibule and at the classroom doors.}
\end{figure*}

We show summary statistics across $100$ simulations in Figure~\ref{fig:baselineBidir}, as well as time snapshots illustrating how students from the two classes interact in an example simulation. Prior to the separation time of $90$~s, only exiting students are active. As students in the entering class begin to arrive at the building doors, we observe more interactions between the two classes, especially in the vestibule and at the classroom doors. Once all of the exiting students have left the building, entering students are the only pedestrians present. In our model implementation, these observations correspond to changes in the set of active students $\mathcal{A}_t$. The mean overlap time in which students from both classes are active (i.e., the time frame in which the mean number of entering students and the mean number of exiting students in the building are both positive) is illustrated in Figure~\ref{fig:baselineBidir}(b).

Since more students are present, mean speeds stay below the levels that our pedestrians achieved in entering- or exiting-only simulations, as we show in Figure~\ref{fig:baselineBidir}(a). These social-force interactions also increase the mean time
that students take to reach their desired desk or building door, depending on their class. The travel-time distributions for entering and exiting students in Figure~\ref{fig:baselineBidir}(c) both show longer tails than the corresponding distributions in Figures~\ref{fig:entering_avgsims}(c) and \ref{fig:exit_avgsims}(c).

\subsection{Effects of Lecture-Hall Size on Classroom-Turnover Times}\label{sec:size}

Larger lecture halls require students to travel farther, so we expect the time that it takes a single class to enter or an isolated class to exit to scale with the size of the room. In cases of bidirectional movement, the possibility for congestion at the classroom doors adds further complexity. Here we investigate the impact of class size on the time that it takes for the prior class to exit and the new one to find their desks. We show how this classroom-turnover time also depends on the separation time between the end of the first class and when entering students start arriving.

To account for lecture halls of different sizes, we adapt our baseline Rock Hall domain to produce four more domains with $200$, $328$, $500$, and $600$ desks, respectively; see Supplementary Material for details. As we show in Supplementary Figure~2, all of our domains have the same vestibule, doors, and aisle widths, and they differ primarily in aisle length, rather than row length. This means that students spend more time moving down aisles as the room size increases, but the distance that they travel in tight rows of desks is about the same across our domains. In all of our simulations, we assume class sizes of $200$ students in our $200$-desk room, $300$ in our $328$-desk room, $400$ in our baseline $416$-desk room (Rock Hall), $500$ in our $500$-desk room, and $600$ in our $600$-desk room; we set $N^\text{early} =0$ and $N^\text{enter} = N^\text{exit}$.

\begin{figure*}
    \centering
    \includegraphics[width=\textwidth]{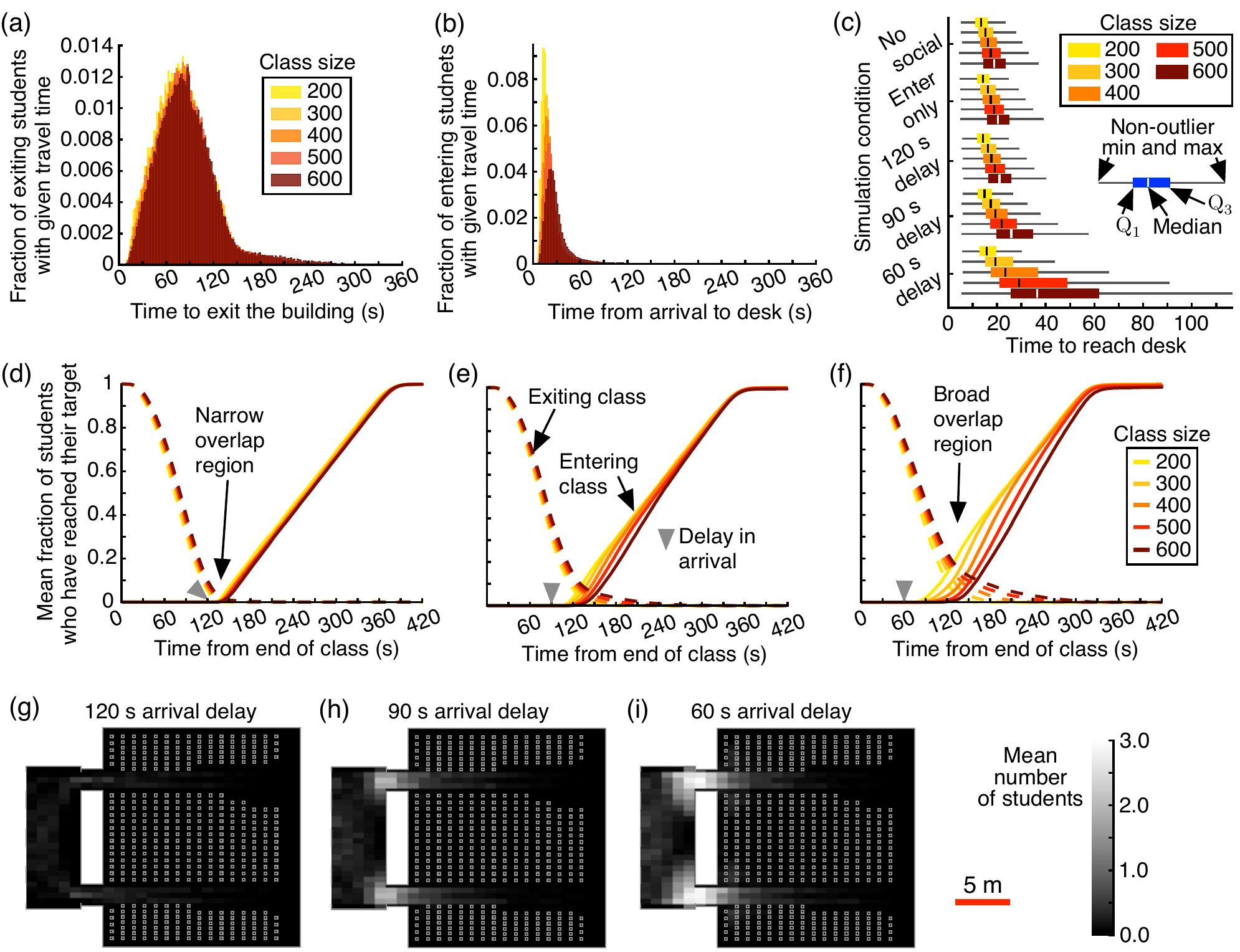}
    \caption{Consecutive classes for different lecture-hall sizes and separation times. We show summary statistics across $100$ simulations for each class size and separation time. (a) The travel-time distributions for exiting students are positive-skewed. As class size increases, the long tails become more prominent. (b) The distributions for the time that it takes entering students to reach their desks (once they arrive at the building doors) also have long tails, and the mean travel time increases with increasing class size. Panels (a)--(b) show different class sizes under a separation time of $90$~s; histogram bin width is $3$~s. (c) Medians, first and third quartiles ($\text{Q}_1$ and $\text{Q}_3$), and non-outlier minimum and maximum travel times for entering students under different conditions. We observe a shift in the median travel time and growing positive skew as class size increases and separation time (or ``delay" after the end of the class before the arrival of entering students) decreases. We define outlier travel times as values that fall below $\text{Q}_1 - 1.5(\text{Q}_3 -\text{Q}_1)$ or above $\text{Q}_3 + 1.5(\text{Q}_3 -\text{Q}_1)$; see Supplementary Material for details. ``No social" refers to setting $B^\text{rep} = B^\text{col} =0$ in Equation~\eqref{eq:baseForce}. We compute the results in panels (a)--(c) across all agents in $100$ simulations. (d)--(f) We show the mean fraction of students in the exiting (dashed line) and entering (solid line) class who have reached their target under a separation time of $120$~s, $90$~s, and $60$~s, respectively, across our $100$ simulations. As the separation time decreases, there is a broader region of overlap with both classes present. (g)--(i) To highlight how congestion at the classroom doors increases with decreasing separation time, we show the mean mass (number of students in grid squares with area $1.5$~m$^2$) at $t=120$~s across $100$ simulations of $400$ entering and $400$ exiting students in our Rock Hall domain. See Supplementary Material for how we compute panels (g)--(i), Supplementary Figure~4 for heat maps in time, and Supplementary Tables I--II for additional summary statistics.}
    \label{fig:bidirSizeDelay}
\end{figure*}

Figure~\ref{fig:bidirSizeDelay}(a) illustrates the effects of lecture-hall size on the time that it takes students in the first class to leave the building, assuming a separation time of $90$~s between the end of the first class and the time that students in the next class begin arriving at the building doors. Similarly, Figure~\ref{fig:bidirSizeDelay}(b) shows the time that entering students spend travelling to their desks after they arrive to the building for rooms of different sizes. The mean travel time for exiting students changes relatively little---by about $15$\%---between the $200$- and $600$-person classrooms. In comparison, the mean travel time for entering students increases from about $16$ s for the $200$-person room to about $35$ s for the $600$-person room, more than doubling. 

We now focus on entering students, since lecture-hall size has the largest impact on their travel times and classroom turnover depends on all of the students in the next class reaching their desks. Because the increase in mean travel time for entering students is partly due to the increasingly long tails in our distributions in Figure~\ref{fig:bidirSizeDelay}(b), it is also useful to consider median travel times in Figure~\ref{fig:bidirSizeDelay}(c). We find a similar dynamic: as class size increases from $200$ to $600$ students for a separation time of $90$ s, the median time that it takes entering students to reach their desks grows from about $15$ s to about $26$ s.

To tease out whether increased student interactions are responsible for the additional time that entering students spend travelling in Figure~\ref{fig:bidirSizeDelay}(b) or if this is simply due to an increase in lecture-hall size, we simulate our model with no social forces. This means $\textbf{f}^\text{col} = \textbf{f}^\text{rep}= 0$ in Equation~\eqref{eq:modelV}. Since each pedestrian in our model acts independently when there are no social forces, this is equivalent to considering only one student moving in the lecture hall. As we show in Figure~\ref{fig:bidirSizeDelay}(c), the median time that it takes entering students to reach their desks differs by only about $5$ s between the $200$- and $600$-person classes, under the case of no social forces. 

If we instead include social forces but assume that there is no exiting class, the median travel time for entering students increases by about $6$ s between the $200$- and $600$-person classes. In comparison, there is a difference of about $11$ s between the median travel times of entering students for $200$- and $600$-person classes when an exiting class is present (under a separation time of $90$~s). We thus conclude that interactions between students in different classes---rather than simply an increase in the physical distance that students need to travel---are responsible for extending classroom-turnover times as lecture-hall size increases. 

These observations suggest that the impact of lecture-hall size on classroom-turnover times may grow as the separation time between the end of the first class and when students in the next class begin arriving decreases. To test this, we simulate $100$ realizations of our model for different lecture-hall sizes under a separation time of $120$ s and $60$ s, respectively. As we show in Figure~\ref{fig:bidirSizeDelay}(c), the impact of lecture-hall size on the time that entering students spend travelling to their desk (after arriving) is most pronounced under a $60$ s separation time. In contrast, a separation time of $120$ s leads to entering students having similar dynamics to the setting where there is no exiting class. Across our simulations, as lecture-hall size increases and separation time decreases, the median travel time for entering students increases and the long tail in our distributions becomes more prominent. 

Our results in Figures~\ref{fig:bidirSizeDelay}(a)--(c) take a student-centered perspective, evaluating how long individuals spend travelling to their desks after they arrive at the building doors. To better understand the influence of separation time on overall classroom dynamics, we next consider the mean fraction of exiting and entering students who have reached their final target (a door or desk) as a function of time from the end of class, across $100$ simulations for each lecture-hall size and separation time. As we show in Figures~\ref{fig:bidirSizeDelay}(d)--(f), the mean fraction of exiting students still in the building decays from one to zero in time, and the mean fraction of entering students who are not yet at their desk, initially zero until the separation time has passed, grows as students arrive at the building doors and later reach their desks.

For the $120$ s separation time in Figure~\ref{fig:bidirSizeDelay}(d), nearly all of the exiting students have left the building before entering students begin arriving. This can also be seen in Figure~\ref{fig:bidirSizeDelay}(g), which provides a heat map of the number of students in the building at $t=120$ s. In comparison, a separation time of $90$ s leads to a longer time interval in which both exiting and entering students are active. As we show in Figure~\ref{fig:bidirSizeDelay}(h), this creates congestion near the classroom doors. Finally, a separation time of $60$ s in Figure~\ref{fig:bidirSizeDelay}(f) produces an even longer overlap region and heavy congestion (Figure~\ref{fig:bidirSizeDelay}(i)) at classroom doors.

Taken together, Figures~\ref{fig:bidirSizeDelay}(a)--(f) highlight that separation times have an opposite effect on the overall classroom dynamics than they do on the experiences of many individual students. As likely intended when scheduling classes, we find that a reduction in the time between classes generally leads to faster overall classroom-turnover times, amongst the separation times that we consider: Figures~\ref{fig:bidirSizeDelay}(d)--(f) show that the time that it takes from the end of the first class for $90$\% of the next class, on average, to have reached their desks is shorter for smaller separation times in our model. On the other hand, most entering students travel for a longer time to reach their desks as the separation time decreases. Our results therefore suggest that shorter separation times may increase frustration for these students, who arrive closer to the end of the first class and must wait in the vestibule due to congestion before they can enter the classroom.

\section{Discussion and Conclusions}

Motivated by large lecture halls at colleges in the United States \cite{RockHall, OhioStateReport, CollegeData}, we developed an off-lattice model for the movement of pedestrians in academic spaces. With a basis in the well-studied social-force approach \cite{helbing1995social,Helbing2009}, our model allows students in the entering class to arrive at building doors, move through a vestibule, and travel down classroom aisles to reach the row of their desk. Similarly, we modeled students in the exiting class as agents who leave their desks at random times after class ends and move toward building doors. We investigated classroom dynamics---with a focus on classroom-turnover times and the travel times of individual students---under different class settings (namely entering only, exiting only, or bidirectional), lecture-hall sizes, and separation times between classes.

Simulating our model without social forces showed that interactions between students play a critical role in classroom-turnover times. The separation time between the end of class and when students in the next class begin arriving determines how much exiting and entering students interact, and shorter separation times increase congestion at classroom doors. We found that the impact of the separation time between classes on student travel times is more pronounced in larger lecture halls. Interestingly, we also observed a discrepancy between the effect that separation time has on the travel time of individuals and on the time that it takes to empty and fill a classroom. While the median travel time for entering students decreased with increasing separation time, the time required to fill $90$\% of the class increased with increasing separation time, particularly in large lecture halls. This suggests that it is important to balance different perspectives when evaluating how the size of a lecture hall will influence student dynamics. 

There are many directions for future work and points of improvement for the model proposed. For example, it is worth noting that the differences in student travel times for different room sizes seem relatively small---on the order of seconds---in Figure~\ref{fig:bidirSizeDelay}. Indeed, while in agreement with prior studies of free, unhindered pedestrian movement \cite{waldau2007pedestrian}, the speed that agents move in our model may be too high for a classroom environment. On the other hand, from an individual pedestrian's perspective, needing to wait an additional $16$~s to enter a $200$- versus a $600$-person room may be tangible and frustrating. Future work could adjust our parameters to match data specific to lecture halls or theaters, and investigate how students perceive different waiting times \cite{Sieben2017}. It is also important to consider a broader range of pedestrian movement, including students travelling with wheelchairs or crutches. The social-force approach can also incorporate additional features such as an effective angle of sight \cite{helbing1995social,helbing2005self} (to model pedestrians in front of a student having a larger impact on their movement than individuals behind them), or attractive forces between students in the same class \cite{helbing1995social} (which may lead to the formation of pedestrian groups). It would also be interesting to consider approaches such as the centrifugal-force model, which modifies the repulsion between two pedestrians depending on if the person in front is slowing down \cite{Yu2005}.

Our results on separation times suggest that individuals may shorten how long it takes to reach their desk by choosing to arrive at the building doors later, particularly for large lecture halls. In contrast, from a collective perspective, arriving later prolongs the time that it takes to fill the classroom. In the future, it may be interesting to incorporate game-theoretic decision making into our model. Using our modeling setup, one could test whether different arrival choices create a dilemma in which the interests of individuals (i.e., minimizing their social interactions and travel times) conflict with the collective interest of emptying and filling the lecture hall in an efficient manner. Similar questions of strategic arrival times have been explored in many contexts, including queuing systems \cite{rapoport2004equilibrium,juneja2013concert}, congestion and congestion pricing in traffic \cite{levinson2005micro,ziegelmeyer2008road}, and start times for meetings \cite{gueant2011mean}. 

More generally, our model does not include dynamic changes in the rules governing pedestrian behavior, and we assign each student their desk and building door at the start of a simulation. Future work could include allowing students to actively select their desks and doors in response to the current conditions in the lecture hall. For example, students could attempt to minimize the number of people that they must pass in narrow rows to reach their desk, or they could seek to optimize their preference to be near the front or the back of the class. It would also be interesting to extend our study to allow students to moderate their speeds based on perceived justness \cite{Sieben2017} or urgency. While our focus has been on understanding the role of lecture-hall size in routine settings, our modeling setup and code \cite{code} could be adapted to investigate other conditions, including evacuation dynamics, lecture-hall design, or---related to the Covid-$19$ pandemic \cite{Harapan2020}---classroom dynamics under social distancing \cite{sajjadi2020social}.

\section*{Appendix A: A-test Analysis} \label{app:uncertainty}

Our model includes several stochastic features, such as each student's initial position and final destination (e.g., desired desks for the entering class). Uncertainty quantification is helpful to determine the number of simulations needed to capture the variance in the summary statistics that we use to describe our results, as shown in \cite{alden2013spartan,cosgrove2015agent,read2012techniques}. We perform this analysis using the A-test measure, as described in~\cite{vargha2000critique}. The A-test involves completing $20$ sets of $k$ simulations each with the same parameters, and extracting a relevant measure for describing the simulations. Each set from $1$ to $20$ is then compared with set $1$ based on this measure to return an A-score, which represents the probability that a randomly selected sample from the first population is larger than a random sample from the second population \cite{cassani2020hybrid}. More specifically, the score $A_{i,j}$ compares sets $i$ and $j$ to determine the significance of the sample size $k$ of simulations within each set. Following the approach in \cite{vargha2000critique}, we compute the discrete approximation of the A-score as:
\begin{equation*}\label{eq:ascore}
A_{i,j} = \dfrac{\# (X_m > Y_n)}{k^2} + 0.5 \dfrac{\# (X_m = Y_n)}{k^2}, 
\end{equation*}
where $ m,n = 1,\ldots,k$ and  $\# (X_m > Y_n)$ is the number of times an element (measure $X_m$) in set $i$ is greater than an element (measure $Y_n$) in set $j$, for all $X_m$'s in set $i$ and all $Y_n$'s in set~$j$. Similarly, $\# (X_m = Y_n)$ is the number of times an element ($X_m$) in set $i$ is equal to an element ($Y_n$) in set $j$. Sets $i$ and $j$ each consist of $k$ simulations.

\begin{figure}
    \centering
    \includegraphics[width=0.48\textwidth]{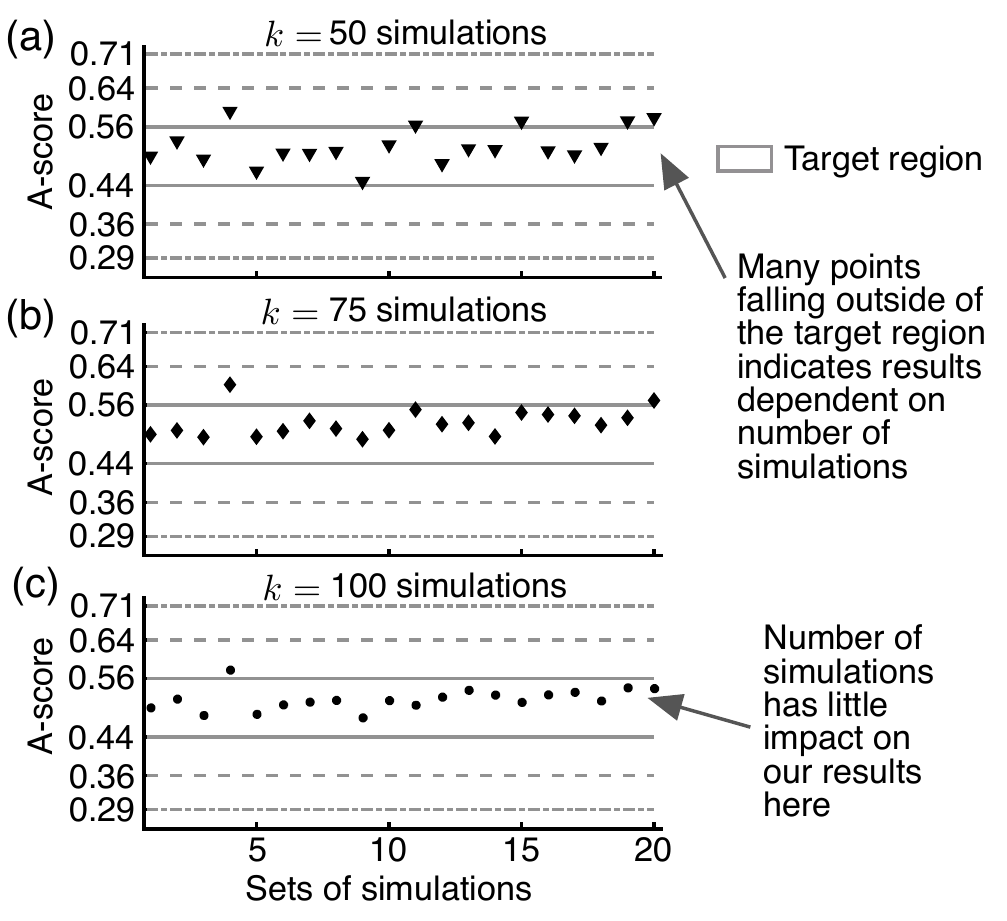}
    \caption{A-test analysis of our model for the case of $400$ students entering our Rock Hall with no exiting class. We show the A-score, computed using student travel times, for (a) $50$, (b) $75$, and (c) $100$ simulations. At $k=100$, all of the A-scores but one fall within the target region (solid lines), so we base our results on $100$ simulations under each condition.
    }
    \label{fig:ascore}
\end{figure}

Figure~\ref{fig:ascore} shows the A-score for $20$ sets of our baseline entering simulations (Figures~\ref{fig:entering_onesim} and \ref{fig:entering_avgsims}), with sample sizes ranging from $k=50$ to $k=100$ and using the parameter values in Table~\ref{tab:parameters}. We use the mean time that it takes students to reach their seat as our measure to compute the A-scores. The results in Figure~\ref{fig:ascore} are based on $k$ simulations of $400$ students entering our Rock Hall. The different lines in Figure~\ref{fig:ascore} provide a means of interpreting test results: scores above $0.71$ or below $0.29$ (outside the dot-dashed lines) mark a large effect of the sample size $k$ on the model results; scores above $0.64$ or below $0.36$ (outside the dotted lines) indicate a medium effect of the sample size $k$ on the results, and scores between $0.44$ and $0.56$ (within the solid lines) illustrate a small effect on the model results. Therefore, the closer the A-score is to $0.5$ for the set comparisons, the smaller the effect that the number of simulations $k$ has on the mean travel time for entering students in our model. An A-score of exactly $0.5$ corresponds to no effect of $k$ on the model results. Note that $A_{1,1}=0.5$, since set $1$ is compared to itself, which provides validation for our A-test analysis. 

As the number of simulations increases from $50$  (Figure~\ref{fig:ascore}(a)) to $100$ (Figure~\ref{fig:ascore}(c)), we find that the number of scores $A_{i,j}$ falling within the solid lines increases. This shows that as the number of simulations $k$ increases, the impact of stochastic factors on our results decreases. In Figure~\ref{fig:ascore}(c), where each set contains $k=100$ simulations, we find that all of the A-scores---except for one---fall within the solid line interval closest to $0.5$. This means that $k=100$ simulations are sufficient to capture most of the variance driven by the stochastic elements in our model. With the exception of the example simulations in Figures~\ref{fig:entering_onesim}, \ref{fig:exit_avgsims}(d)--(h), \ref{fig:baselineBidir}(d)--(h) and Supplementary Figure~3, all of our results are thus based on summary statistics calculated across $100$ simulations.

\section*{Appendix B: Model Parameters}\label{app:parameters}

We include all of our model parameters in Table~\ref{tab:parameters}. To support the reproducibility of our work, we also list the simulation-specific values of $N^\text{enter}$, $N^\text{exit}$, $N^\text{early}$, and $t_\text{sep}$ by figure in the Supplementary Material.

\begin{table*}[t!]
\setlength\tabcolsep{6pt}
    \center
    \caption{\label{tab:parameters}Summary of our model parameters. See Supplementary Material for values that are simulation-specific by figure.}
    \begin{tabular}{l p{2.75cm} p{4.6cm}  p{7.2cm}}
    \textbf{Parameter} & \textbf{Value} & \textbf{Source/Motivation} & \textbf{Meaning}  \\
    \hline 
    $v_\text{avg}$    &  $1.34$ m/s &
    Based on \cite{helbing1995social,waldau2007pedestrian} & Mean speed at which students desire to move \\
    $v_\text{std}$    &   $0.37$ m/s &  Based on \cite{waldau2007pedestrian} & Standard deviation for the students' desired speed \\
    $\tau$  &  $1.0$ s&  Based on \cite{helbing2005self} & Relaxation timescale for the force that students outside of desk rows feel to align with their desired velocity (Equation~\eqref{eq:desForce} and Figure~\ref{fig:modelSchematic}(f)) \\
    $\tau_\text{row}$  &  $0.1$ s& Less than $\tau$ to allow for quick changes in velocity in desk rows& Relaxation timescale for the force that aligns students in desk rows with their desired velocity (Equation~\eqref{eq:desForce} and Figure~\ref{fig:modelSchematic}(f))\\
    $\delta_t$  &  $0.1$ s& Similar to choice in \cite{Helbing2009} to reduce student collisions  &  Timescale for how far in advance students can predict the movement of others (Equations~\eqref{eq:ellipticaldistance}--(9))  \\
    $B^\text{rep}$ &   $0.11$  m/s$^2$ & Based on \cite{johansson2007specification} & Repulsion strength to maintain a comfortable distance between pedestrians (Equation~\eqref{eq:baseForce} and Figure~\ref{fig:modelSchematic}(e))\\ 
    $b^\text{rep}$ &  $0.84$ m   & Based on \cite{johansson2007specification} & Interaction range for repulsive force between students (Equation~\eqref{eq:baseForce} and Figure~\ref{fig:modelSchematic}(e)) \\ 
   $B^\text{col}$ &   $0.11$  m/s$^2$ & Chosen equal to $B^\text{rep}$ &Repulsion strength in the collision-avoidance force between students (Equation~\eqref{eq:baseForce} and Figure~\ref{fig:modelSchematic}(e)) \\
   $b ^\text{col}$ &  $0.084$ m   & Chosen to be $0.1 B^\text{rep}$ & Interaction range for very local force to prevent collisions (Equation~\eqref{eq:baseForce} and Figure~\ref{fig:modelSchematic}(e)) \\ 
    $r$   & $0.6$ m    & Based on \cite{helbing2005self} & Diameter of the privacy sphere that pedestrians strive to maintain (Equation~\eqref{eq:baseForce})\\
    $\sigma$    &$0.001$ m/s$^{3/2}$  & Chosen to include weak noise in pedestrian movement &Noise strength in the differential equation for student velocity (Equation~\eqref{eq:modelV})\\
    $\mu$ & $0.01$~m & Helps prevent rare head-on collisions in the vestibule& Small amount of noise in the classroom destination $\textbf{D}^\text{class}_i$ for entering student $i$ (Equation~(S3) in Supplementary Material) \\
    $N^\text{enter}$ & Simulation-specific & Accounts for different class sizes & Number of students in the entering class\\
    $N^\text{early}$ & Simulation-specific&Includes a small number of early students in some simulations & Number of early-arriving students who are in $\mathcal{A}_0$ and start in the vestibule\\
    $N^\text{exit}$ & Simulation-specific&Accounts for different class sizes & Number of students in the exiting class \\
    $t_\text{sep}$ & Simulation-specific & Range of separation times tested & Time in seconds between when the first class ends and when entering students begin arriving \\ 
    $d_\text{tol}$   &$0.3$ m& Chosen to be $r/2$  & Tolerance distance when updating row status and $t^\text{final}$ (Equations~\eqref{eq:finalTime}--\eqref{eq:travTime})\\ 
    $b_\text{bnd}$ &$0.6$ m & Chosen to be $r$ & Distance from the building walls in which velocity is reduced for boundary conditions (Equation~(S7) in Supplementary Material)\\
    $b_\text{tight}$ & $0.3$ m  & Less than $b_\text{bnd}$ to accommodate movement in tight spaces & Distance from doors, aisle walls, or desk-row walls in which velocity is reduced for boundary conditions (Equation~(S7) in Supplementary Material) \\
    $\alpha$ & $0.004175 N^\text{enter}$ students/s  & Chosen so that the entering class arrives over about $4$ minutes 
    & Rate of entering students arriving at the building doors (Equation~\eqref{eq:probActive})\\
    $\mu_\mathrm{pre}$ & $60$ s & Informed roughly by evacuation experiments in \cite{nilsson2009social} & Mean pre-movement time for exiting students \\
    $\sigma_\mathrm{pre}$ & $35$ s & Informed roughly by evacuation experiments in \cite{nilsson2009social} & Standard deviation in the pre-movement times for exiting students\\
    $\lambda^\mathrm{pre}_\mathrm{min}$ & $0$ s & Chosen so exiting students do not leave before class ends& Cutoff parameter in our distribution for $\tau^\text{pre}_i$ \\ 
    $\lambda^\mathrm{pre}_\mathrm{max}$ & $120$ s & Chosen to be $2\mu_\text{pre}$& Cutoff parameter in our distribution for $\tau^\text{pre}_i$ \\ 
    \hline
    \end{tabular}
\end{table*}

\section*{Acknowledgments}
This collaboration is based upon work supported by the Mathematics Research Communities of the American Mathematical Society (AMS), under National Science Foundation (NSF) grant no.\ DMS-1641020. Our project was initiated during the 2018 MRC on Agent-based Modeling in Biological and Social Systems. We are also grateful to the AMS MRC program for supporting a follow-up collaboration visit to the Mathematical Biosciences Institute. This collaboration was also supported by the NSF under grant no.\ DMS-1929284 through a visit to the Institute for Computational and Experimental Research in Mathematics in Providence, RI, during the ``Mathematical Models of Pedestrian Movement in Large Lecture Halls" Collaborate@ICERM program. We thank  Danielle Ciesielski and Andrew Bernoff for helpful discussions at the MRC workshop, and we also thank Chad Topaz for connecting AV and KB.

\bibliographystyle{unsrt}  
\bibliography{references} 

\appendix

\renewcommand{\thesection}{SI \arabic{section}} 

\renewcommand{\thefigure}{SI~\arabic{figure}}
\setcounter{figure}{0}
\renewcommand{\thetable}{SI~\arabic{table}}
\setcounter{table}{0}
\renewcommand{\theequation}{SI~\arabic{equation}}
\setcounter{equation}{0}

\section{Supplementary Material}

To support the reproducibility of our results, this Supplementary Material includes:
\begin{itemize} \vspace{-0.5\baselineskip}
\item information about the lecture-hall domains in which we simulate student movement (Section \ref{app:lecture});\vspace{-0.5\baselineskip}
\item details about the initial conditions and student destinations in our model (Section \ref{app:destination});\vspace{-0.5\baselineskip}
\item details about how we implement boundary conditions in our model (Section \ref{app:boundaries}); \vspace{-0.5\baselineskip}
\item information about the numerical implementation of our model (Section \ref{app:numerical}); \vspace{-0.5\baselineskip}
\item a summary of the figure-specific values of our model parameters and details on how we compute heat-map images (Section \ref{app:parameters}); and \vspace{-0.5\baselineskip}
\item additional simulation figures, including timelines of student density for different separation times between classes and an example agent trajectory for an outlier in our simulations (Section \ref{app:simulation}).
\end{itemize}\vspace{-0.5\baselineskip}
The code that we developed to simulate our model will be made available at \cite{code}.

\subsection{Additional Modeling and Implementation Details}\label{app:modelDetails}

We provide additional details on our lecture-hall domains, initial conditions and student destinations, boundary conditions, and numerical implementation, as well as information on how to reproduce our figures, in Sections \ref{app:lecture}--\ref{app:parameters}, respectively.

\subsubsection{Lecture-Hall Domains}\label{app:lecture}

As we show in Supplementary Figure~\ref{fig:motivation}, lecture halls that can accommodate $300$ or more students are not uncommon in large colleges and universities in the Unites States. Motivated by the broad range of classroom sizes in Supplementary Figure~\ref{fig:motivation}(b), we thus construct five domains, pictured in Supplementary Figure~\ref{fig:rooms}. Our baseline domain is modeled after our measurements of Rock Hall \cite{RockHall} at the University of California, Davis (see Supplementary Figures~\ref{fig:motivation}(c)--(d)) and has $416$ desks. In this domain, which is meant to be a simplified approximation of Rock Hall, the average distance from each desk to its nearest-neighboring desk is $0.543$ m, measured between desk centers, and we do not directly account for the presence of stairs in Supplementary Figures~\ref{fig:motivation}(c)--(d). We also consider two smaller classrooms that hold $200$ and $328$ desks, respectively, and two larger lecture halls with $500$ and $600$ desks, respectively. Our $328$-desk room models an academic space with a capacity that is in between two other large classrooms at the University of California, Davis \cite{DavisClassrooms} with capacity for $369$ and $285$ people, respectively. Motivated by our measurements of Rock Hall, we assume building doors are $1.8$ m wide, internal classroom doors are $1.75$ m wide, and classroom aisles are $2.0$ m wide. 

Because our goal is to investigate the role of lecture-hall size, rather than room geometry, in classroom-turnover times, we account for different room sizes by adjusting the length (and to a much lesser extent, the width) of the classroom only. Our $600$-person domain has a classroom length of $27$~m (not including the vestibule) and width of $20$~m; similarly, our $500$-person room is $23$~m long and $20$~m wide, our $416$-person room is $20$~m long and $20$~m wide, our $328$-person room is $17$~m long and $20$~m wide, and our $200$-person room is $12$~m long and $19$~m wide.  We use the same vestibule geometry (namely $5$~m long and $13$~m wide) and door sizes for all of our domains; see Supplementary Figure~\ref{fig:rooms}. See our code to be posted on GitLab \cite{code} for the coordinates of desk centers, doors, and aisles in our domains.

\begin{figure*}
\includegraphics[width=\textwidth]{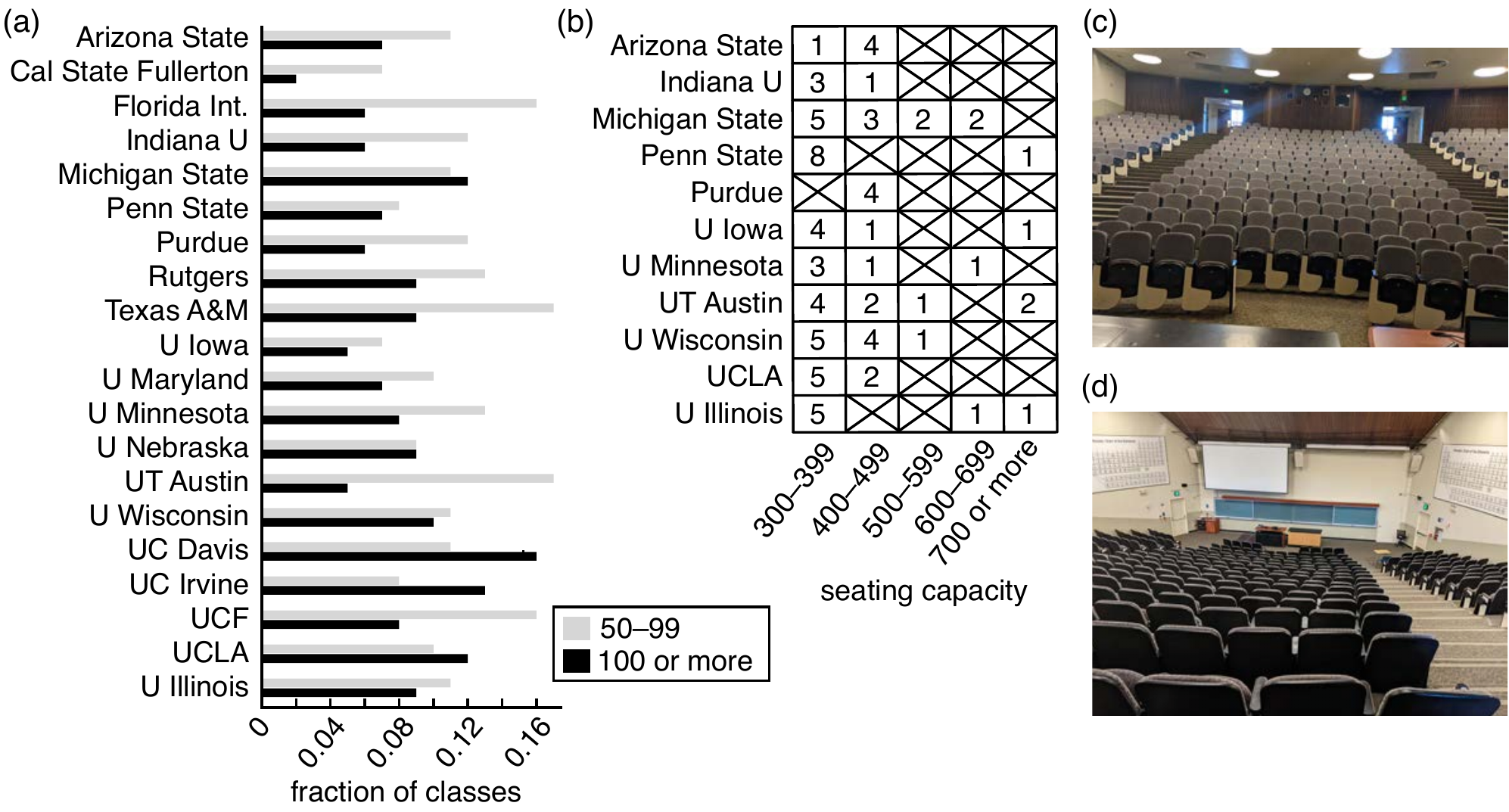}
\caption{Overview of classroom sizes at example universities and colleges in the United States. (a) Across the twenty institutions that we show (namely examples of public, Big Ten Academic Alliance and large institutions with data available here \cite{CollegeData}), on average about $8$\% of classes have $100$ or more students and about $12$\% have $50$--$99$ students. Importantly, these percentages are in terms of classes, rather than students; this means that the percentage of students who find themselves in a large class is much higher. (Our estimates are unweighted averages and based on data from \emph{CollegeData.com} \cite{CollegeData}.) (b) A 2009 report associated with The Ohio State University highlighted the number of large lecture halls by size at different universities, and we show a sample of this data \cite{OhioStateReport}. (c)--(d) As an example academic space, Rock Hall \cite{RockHall} at the University of California, Davis has a capacity for $416$ people.}
\label{fig:motivation}
\end{figure*}

\begin{figure*}
    \centering
    \includegraphics[width=\textwidth]{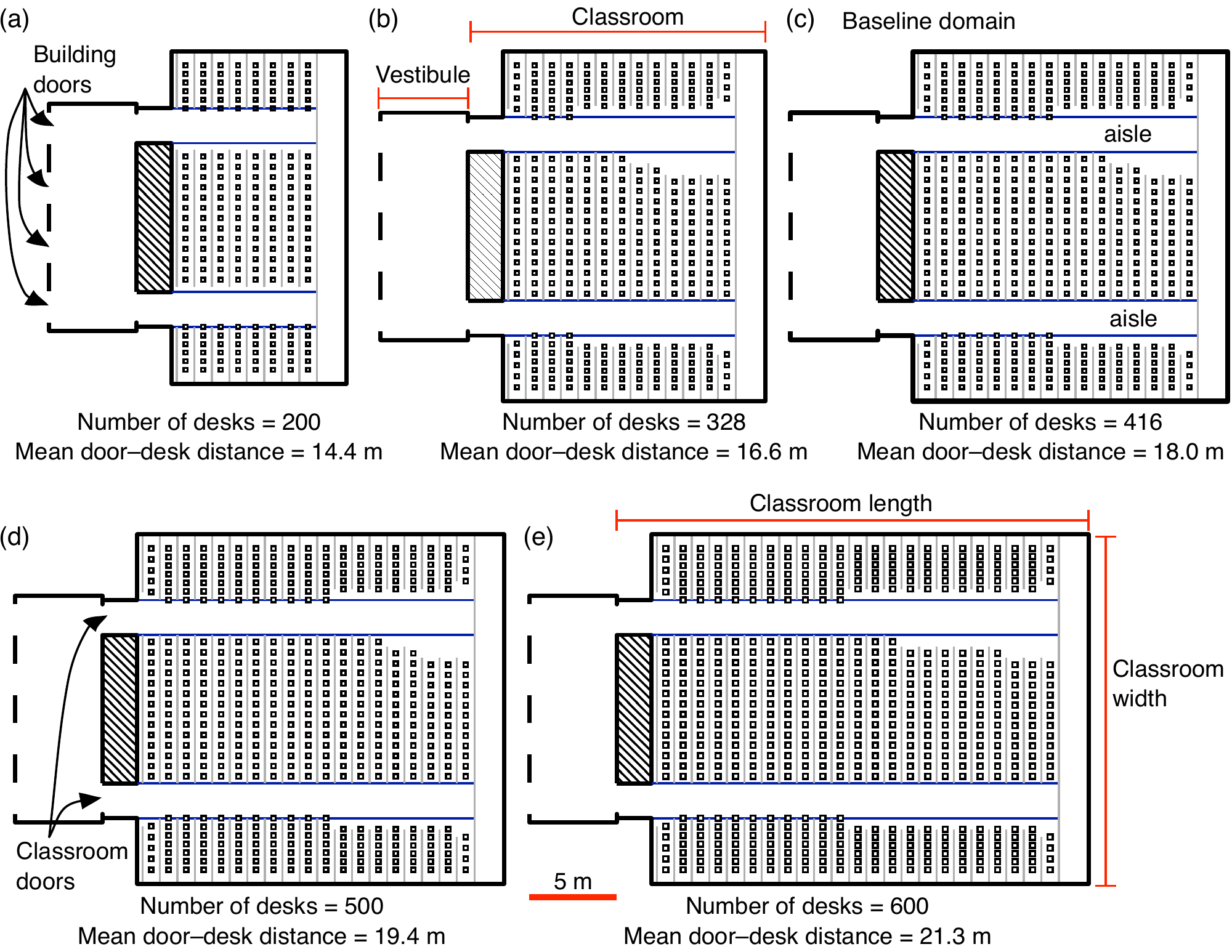}
    \caption{Overview of our lecture-hall domains. All of these domains are based on adapting our rough estimate of Rock Hall at the University of California, Davis. The same scale bar applies to panels (a)--(e). We indicate aisle boundaries in blue, desk-row boundaries in gray, and desks as squares; see Figure~1(a) in the main manuscript for details. (a) Our smallest lecture hall contains $200$ desks, with an average distance of $0.5405$ m between each desk center and its nearest-neighboring desk center. (b) Our next classroom has $328$ desks, and the average distance between nearest desk centers is $0.5419$~m. (c) Our baseline domain, approximated from our measurements of Rock Hall, contains $416$ desks with an average separation of $0.543$ m between desks. (d) Our second largest domain has $500$ desks separated by an average distance of $0.5428$ m, and (e) our largest lecture hall contains $600$ desks with an average separation of $0.5405$ m between nearest-neighboring desks. We refer to the horizontal length of the classroom space in our domains as the ``classroom length" and the vertical height as the ``classroom width". The vestibule is the same size in all of our domains, and we include four building doors (between the vestibule and outside), two internal classroom doors (between the vestibule and classroom), and two aisles in all cases.}
    \label{fig:rooms}
\end{figure*}

\subsubsection{Assigning Student Destinations and Initial Conditions}\label{app:destination}

As we overview in Section II B of the main manuscript, we assign each student $i$ a desk with coordinates $\textbf{D}^\text{desk}_i$ and a building door with coordinates $\textbf{D}^\text{bldg}_i$; we also assign them two intermediate destinations, a classroom door $\textbf{D}^\text{class}_i$ and an aisle destination $\textbf{D}^\text{aisle}_i$. We follow the process below:
\begin{align}
    \textbf{D}^\text{desk}_{i} &= \textbf{d}_j \text{     with $j \sim \text{Unif}^\text{discrete}(1,N^\text{desks})$ (student $i$'s desk is chosen uniformly at random from the desks)},\nonumber \\
   \textbf{D}^\text{bldg}_{i} &= \textbf{b}_k + (0.5, \beta_i) \text{     with $k \sim \text{Unif}^\text{discrete}(1,4)$ ($i$ chooses their door $k$ uniformly at random)}, \label{eq:bldgDest}\\
    \textbf{D}^\text{aisle}_{i} &= \begin{cases}
    ({D}^\text{desk}_{i,1}, a_\text{up}) & \text{if $|{D}^\text{desk}_{i,2} - a_\text{up}| < |{D}^\text{desk}_{i,2} - a_\text{low}|$ ($i$ uses the upper aisle if it is closer to their desk)}  \\
    ({D}^\text{desk}_{i,1}, a_\text{low}) & \text{otherwise ($i$ uses the lower aisle if it is closer to their desk)} 
    \end{cases}, \label{eq:aisleDest} \\
    \textbf{D}^\text{class}_{i} &= \begin{cases}
    \textbf{c}_\text{up} + \textbf{w}_i & \text{if $|{D}^\text{desk}_{i,2} - a_\text{up}| < |{D}^\text{desk}_{i,2} - a_\text{low}|$ ($i$ chooses the classroom door associated with their $\textbf{D}^\text{desk}_{i}$)} \\
    \textbf{c}_\text{low} + \textbf{w}_i & \text{otherwise ($i$ chooses the classroom door associated with their $\textbf{D}^\text{desk}_{i}$)} 
    \end{cases},\label{eq:classDest}
\end{align}
where $\textbf{d}_j$ marks the $(x,y)$-coordinates of the center of the $j$th desk; $\text{Unif}^\text{discrete}(1,n)$ is the discrete uniform distribution from $1$ to $n$; $N^\text{desks}$ is the number of desks in the lecture hall; $\textbf{b}_k$ is the midpoint of the $k$th door into the building, and there are $n=4$ building doors in each of our domains; $\textbf{D}^\text{desk}_i = (D^\text{desk}_{i,1},D^\text{desk}_{i,2})$ are the coordinates of student $i$'s desk; $a_\text{up}$ (respectively, $a_\text{low}$) is the $y$-coordinate of the center of the upper (respectively, lower) aisle (see Figure~1(a) in the main manuscript); $\textbf{c}_\text{up}$ (respectively, $\textbf{c}_\text{low}$) marks the midpoint of the classroom door that leads into the upper (respectively, lower) aisle of the classroom; $\textbf{w}_i = (w_{i,1},w_{i,2})$ with $w_{i,k} \sim \text{Unif}(0,\mu)$ for $k \in \{1,2\}$; and $\beta_i \sim \text{Unif}(-0.5,0.5)$ m. We use $\mu = 0.01$~m for students in the entering class, and $\mu = 0$~m for exiting students. 

For each student $i$ in the entering class, we include a small amount of noise in $\textbf{D}^\text{class}_{i}$ to help prevent rare occasions where we observed pairs of agents getting stuck along the vestibule--classroom wall early on in our model-development process. In this rare setting, particularly in crowded conditions, some agents were pushed against the vestibule--classroom wall and moving toward one another in order to reach their classroom destinations; by adding some noise to  $\textbf{D}^\text{class}_{i}$, we help prevent these agents from getting stuck at a head-to-head standstill in their attempt to move in directly opposing directions. Additionally, we add $0.5$ m to the $x$-coordinate of $\textbf{b}_k$ when computing $\textbf{D}^\text{bldg}_{i}$ in Equation~\eqref{eq:aisleDest} to ensure new entering students are inside the building boundary. While this choice is motivated by entering students, we add $0.5$ m to the $x$-coordinate of $\textbf{D}^\text{bldg}_k$ for both exiting and entering students so that students in both classes have the same distance to travel in our bidirectional simulations. 

Each student $i$ in the exiting class begins active at their desk, so $\textbf{X}_i(0) = \textbf{D}^\text{desk}_i$. In our simulations with only an entering class, we assume that $2$\% of the entering class starts in the vestibule. This means that we include $N^\text{early} = 0.02 N^\text{enter}$ early-arriving students in the entering class when $N^\text{exit} =0$; these early-arriving students have a starting point in the vestibule, modeling students waiting outside of the classroom. Specifically, if student $i$ is an early-arriving student, then they are in $\mathcal{A}_0$, and we set $\textbf{X}_{i}(0)$ to be a position selected uniformly at random from a grid of points in the vestibule (with a step size of $1$ m, specified to ensure that students are not randomly initialized at locations that are unrealistically close). With the exception of these $N^\text{early} \ge 0$ early-arriving students, all of the other students in the entering class start at a building door when they become active; in particular, if student $i$ becomes active at time $t^\text{active}_i$, then $\textbf{X}(t^\text{active}_i) = \textbf{D}^\text{bldg}_i$. In Equation~\eqref{eq:bldgDest}, if $N^\text{enter} - N^\text{early}$ (respectively, $N^\text{exit}$) is not divisible by four (the number of building doors), we use a distribution that is approximately uniform to assign each student a building door $k$ from the set of all doors.

\subsubsection{Boundary Conditions}\label{app:boundaries}

For each student $i \in \mathcal{A}_t$, after calculating $\textbf{V}_i(t+\Delta t)$ according to Equation~(4) in the main manuscript, we adjust $\textbf{V}_i(t+\Delta t)$ as necessary to account for no-flux boundary conditions along the building walls, row boundaries, and---depending on student $i$'s row status $R_i(t)$---aisle boundaries (see Figures~1 and 2 in the main manuscript). First, we note that our building outline can be represented by a set of line segments, and all of the walls that we consider are straight and parallel to either the $x$- or $y$-axis. For each student $i\in\mathcal{A}_t$, we find the point $\textbf{b}^\text{bldg}_i$ on the building-boundary line segments that is closest to their current position $\textbf{X}_i(t)$ and compute the unit vector $\textbf{e}^\text{bldg}_i = (\textbf{b}^\text{bldg}_i - \textbf{X}_i(t))/||\textbf{b}^\text{bldg}_i - \textbf{X}_i(t)||$. Second, since the boundaries of all of the aisles and rows are also line segments that are parallel to the $x$- and $y$-axes, respectively, we follow a similar process for the desk rows and aisles (Figure~1(a) in the main manuscript). In particular, we find the point $\textbf{b}^\text{r}_i$ on the desk-row boundaries that is closest to $\textbf{X}_i(t)$ and compute the unit vector $\textbf{e}^\text{r}_i = (\textbf{b}^\text{r}_i - \textbf{X}_i(t))/||\textbf{b}^\text{r}_i - \textbf{X}_i(t)||$. If $R_i(t) = 0$, we also find the point $\textbf{b}^\text{a}_i$ on the aisle boundaries that is closest to $\textbf{X}_i(t)$ and calculate $\textbf{e}^\text{a}_i = (\textbf{b}^\text{a}_i - \textbf{X}_i(t))/||\textbf{b}^\text{a}_i - \textbf{X}_i(t)||$. 

Third, we compute student $i$'s updated velocity, $\textbf{V}^{**}_i$, which accounts for our no-flux boundary conditions, in a three-step process, as below:
\begin{align}
\textbf{v}_i &= \textbf{V}_i(t+\Delta t) \underbrace{-\bold{f}^\text{bnd}(\textbf{X}_i(t),\textbf{V}_i(t+\Delta t), \textbf{e}^\text{bldg}_i,\textbf{b}^\text{bldg}_i)}_\text{accounts for the building walls} \label{eq:bc1}\\
\textbf{v}^{*}_i &= \begin{cases}
\textbf{v}_i- \underbrace{\bold{f}^\text{bnd}(\textbf{X}_i(t),\textbf{v}_i, \textbf{e}^\text{a}_i,\textbf{b}^\text{a}_i)}_\text{accounts for the aisles} & \text{    if $R_i(t) = 0$}\\
\textbf{v}_i & \text{    otherwise}
\end{cases}\label{eq:bc2} \\
\textbf{v}^{**}_i &= \textbf{v}^*_i- \underbrace{\bold{f}^\text{bnd}(\textbf{X}_i(t),\textbf{v}^*_i, \textbf{e}^\text{r}_i,\textbf{b}^\text{r}_i)}_\text{accounts for the desk rows}  \label{eq:bc3}
\end{align}
where $\bold{f}^\text{bnd}(\textbf{x},\textbf{v},\textbf{e},\textbf{b}) = f^\text{bnd}(\textbf{x},\textbf{b})\text{proj}_{\textbf{e}}(\textbf{v})$ when both $||\textbf{x}-\textbf{b}|| \le 2b^\text{bnd}$ and $\text{proj}_{\textbf{e}}(\textbf{v}) \ge 0$, and $\bold{f}^\text{bnd} = 0$ otherwise. Here the factor $f^\text{bnd}(\textbf{x},\textbf{b})$ serves as a smoothed heaviside function, so that students who are near a boundary and moving toward it experience a gradual decrease in their speed. This factor is given by:

\begin{align}
    f^\text{bnd}(\textbf{x},\textbf{b}) &= \begin{cases}
    \frac{1}{2} + \frac{1}{2}\text{tanh}\left(10(b^\text{bnd} - ||\textbf{b}-\textbf{x}||)\right) &\text{if $\textbf{b}$ is on the building boundary and is not a door frame}\\
    \frac{1}{2} + \frac{1}{2}\text{tanh}\left(10(b^\text{tight} - ||\textbf{b}-\textbf{x}||)\right) &\text{if $\textbf{b}$ is on an aisle, row, or door-frame boundary}.
    \end{cases}\label{eq:bcFreeze}
\end{align}

We choose $b^\text{tight} < b^\text{bnd}$ to account for student movement in the tighter spaces of door frames, aisles, or rows. Finally, with a slight abuse of notation, we set $\textbf{V}_i(t+\Delta t) = \textbf{v}^{**}_i$.

\subsubsection{Numerical Implementation}\label{app:numerical}

We simulate our model using MATLAB, and our code will be made available on GitLab \cite{code}. As we summarize in Figure~2 in the main manuscript, at each time step $\Delta t = 0.01$~s, we synchronously calculate the updated velocity for each student, adjust these velocities to account for the appropriate no-flux boundary conditions, and update each student's position. We then synchronously update the row status $R_i$ and current destination $\textbf{D}_i$ of each student $i$; check whether any exiting students have left the building and, if so, remove them from the active set; and add any new arriving students to the simulation. We use the ``inpolygon" function in MATLAB to determine whether students are inside the classroom and whether exiting students have left the building in Equations~(4)--(5) in the main manuscript. We implement our stochastic differential equation for student velocity in Equation~(4) in the main manuscript using the Euler--Maruyama method \cite{Higham}; and we implement our differential equation for student position in Equation~(5) in the main manuscript using a forward Euler scheme. We simulate from time $t = 0$ to $t = t_\text{max}$ with a time step of $\Delta t = 0.01$~s.

\subsubsection{Instructions for Reproducing our Figures}\label{app:parameters}

Our model parameters are in Table~II in the main manuscript. Because we investigate different classroom sizes and separation times, we use simulation-specific values of $N^\text{enter}$, $N^\text{exit}$, $N^\text{early}$, $t_\text{max}$, and $t_\text{sep}$. To support reproducibility of our work, we include these parameters, as well as details on our heat-map images, by figure below:
\begin{itemize}
 \item \textbf{Heat-map Images:} The heat maps in Figures~4(d)--(h) and 7(g)--(i) in the main manuscript and Supplementary Figure~\ref{fig:DelayBidirDensity} are based on $100$ simulations for each parameter condition, and here we describe our process for creating each of these figures. At the simulation time point of interest, we discretize our domain from $x=0$ m to $x= 25$~m with $\Delta x = 1$ m and from $y =0 $ m to $y= 20$ m with $\Delta y = 0.5$ m. For each of our $100$ simulations, we find the number of students in each grid square at the given time and compute the mean velocity across the students in that grid square. To produce heat maps of mean student speed, we take the mean of these values across $100$ simulations, omitting the velocity in cases where there are no students in a given grid square. Each grid square contains at most one desk. Because our grid step and desk positions do not line up perfectly, there are grid squares in our classroom in which no desks are located (e.g., see the narrow horizontal strips in the center desk region that appear as black in Figure 4(h) in the main manuscript). 
    \item \textbf{Figure 3:} $N^\text{enter} = 400$ students, $N^\text{exit} = 0$ students, $N^\text{early} = 0.02 \times 400 = 8$ students, $t_\text{max} = 300$ s, and $t_\text{sep} = 0$ s; \vspace{-0.5\baselineskip}
    \item \textbf{Figure 4:} $N^\text{enter} = 400$ students, $N^\text{exit} = 0$ students, $N^\text{early} = 0.02 \times 400 = 8$ students, $t_\text{max} = 280$ s, and $t_\text{sep} = 0$ s; \vspace{-0.5\baselineskip}
    \item \textbf{Figure 5:} $N^\text{enter}=N^\text{early} =0$ students, $N^\text{exit} =400$ students, $t_\text{max} = 280$ s, and $t_\text{sep} = 0$ s;\vspace{-0.5\baselineskip}
    \item \textbf{Figures~6 and 7(h) and Supplementary Figures~\ref{fig:exampleStuck} and \ref{fig:DelayBidirDensity}(i)--(p):} $N^\text{enter} = N^\text{exit} = 400$~students, $N^\text{early} = 0$ students, $t_\text{max} = 450$ s, and $t_\text{sep} = 90$ s;\vspace{-0.5\baselineskip}
    \item \textbf{Figures~7(a)--(b) and 7(e):} $N^\text{enter}=N^\text{exit}$ is the class size in the figure legend, $N^\text{early} = 0$ students, $t_\text{max} =450$~s, and $t_\text{sep} = 90$ s.\vspace{-0.5\baselineskip}
    \item \textbf{Figure~7(c):} Conditions are the same as in Figures~7(d)--(f), as described above and below, for our bidirectional simulations. Under ``No social", we set $B^\text{col} = B^\text{rep} = 0$ m/s$^2$; include $N^\text{exit} = 0$ students; provide $N^\text{enter}$ as the class size in the legend; include $N^\text{early} = 0.02 N^\text{enter}$ students; set $t_\text{max} = 400$ s for the $200$-person class, $t_\text{max}= 100$~s for the $300$-person class with $\alpha = 10.67$ students/s, $t_\text{max} = 600$~s for the $400$- and $500$-person classes (in some cases we simulate for much longer than necessary), and $t_\text{max} = 50$~s with $\alpha = 200.67$ students/s for the $600$-person class; and use $t_\text{sep} = 0$ s. Under ``Enter only", we include $N^\text{exit} = 0$ students; provide $N^\text{enter}$ as the class size in the legend; include $N^\text{early} = 0.02 N^\text{enter}$ students; set $t_\text{max} = 300$ s for the $200$- and $300$-person classes, $280$~s for the $400$-person class, $550$~s for the $500$-person class, and $320$~s for $600$-person class; and use $t_\text{sep} = 0$ s. \vspace{-0.5\baselineskip}
    \item \textbf{Figure~7(d):} $N^\text{enter}=N^\text{exit}$ is the class size in the figure legend; $N^\text{early} = 0$ students; $t_\text{max} =450$~s for the $600$- and $400$--person classes and $500$~s for the  $500$-, $300$-, and $200$-person classes; and $t_\text{sep} = 120$ seconds.\vspace{-0.5\baselineskip}
    \item \textbf{Figure~7(f):} $N^\text{enter}=N^\text{exit}$ is the class size in the figure legend; $N^\text{early} = 0$ students; $t_\text{max} =450$~s for the $600$- and $500$-person classes, $410$~s for the $400$--person class, $500$~s for the $300$-person class, and $600$~s for the $200$-person class; and $t_\text{sep} = 60$ s.\vspace{-0.5\baselineskip}
   \item \textbf{Figure~7(g) and Supplementary Figures~\ref{fig:DelayBidirDensity}(q)--(x):} $N^\text{enter}=N^\text{exit} = 400$ students, $N^\text{early} =0$ students, $t_\text{max} = 450$~s, and $t_\text{sep} = 120$~s.\vspace{-0.5\baselineskip}
    \item \textbf{Figure~7(i) and Supplementary Figures~\ref{fig:DelayBidirDensity}(a)--(h):} $N^\text{enter}=N^\text{exit} = 400$ students, $N^\text{early} =0$ students, $t_\text{max} = 410$~s, and $t_\text{sep} = 60$~s.
\end{itemize}

\subsection{Additional Simulation Examples}\label{app:simulation}

In Supplementary Figure~\ref{fig:exampleStuck}, we highlight an example of a simulation outlier. In particular, we provide the trajectory of an example agent who takes longer than $\text{Q}_3 + 1.5(\text{Q}_3 - \text{Q}_1)$ to reach their desk, for the case of a $600$-person room with a separation time of $60$~s. Here $\text{Q}_3$ is the $75$th percentile of $\{t^\text{final}_i\}_{i \text{ in the entering class}}$ and $\text{Q}_1$ is the $25$th percentile of $\{t^\text{final}_i\}_{i \text{ in the entering class}}$, as determined using the ``prctile" function in MATLAB. The student shown spends about $120$~s in the vestibule, attempting to enter the classroom; note that we omit plotting the other students that are moving around this outlier agent. Once in the classroom, they are pushed into the wrong desk row by other agents. At this point, near $t = 180$~s in Supplementary Figure~\ref{fig:exampleStuck}(b), the student's row status is still zero because they are not in their desired row; thus, they still experience repulsive forces from the aisle walls, in addition to building walls, desk rows, and other students. As a result, this agent spends about a minute in the wrong row, before attractive forces toward their desired aisle destination overwhelm the repulsive forces that the agent experiences from the aisle walls and other students. 

\begin{figure*}
    \centering
    \includegraphics[width=\textwidth]{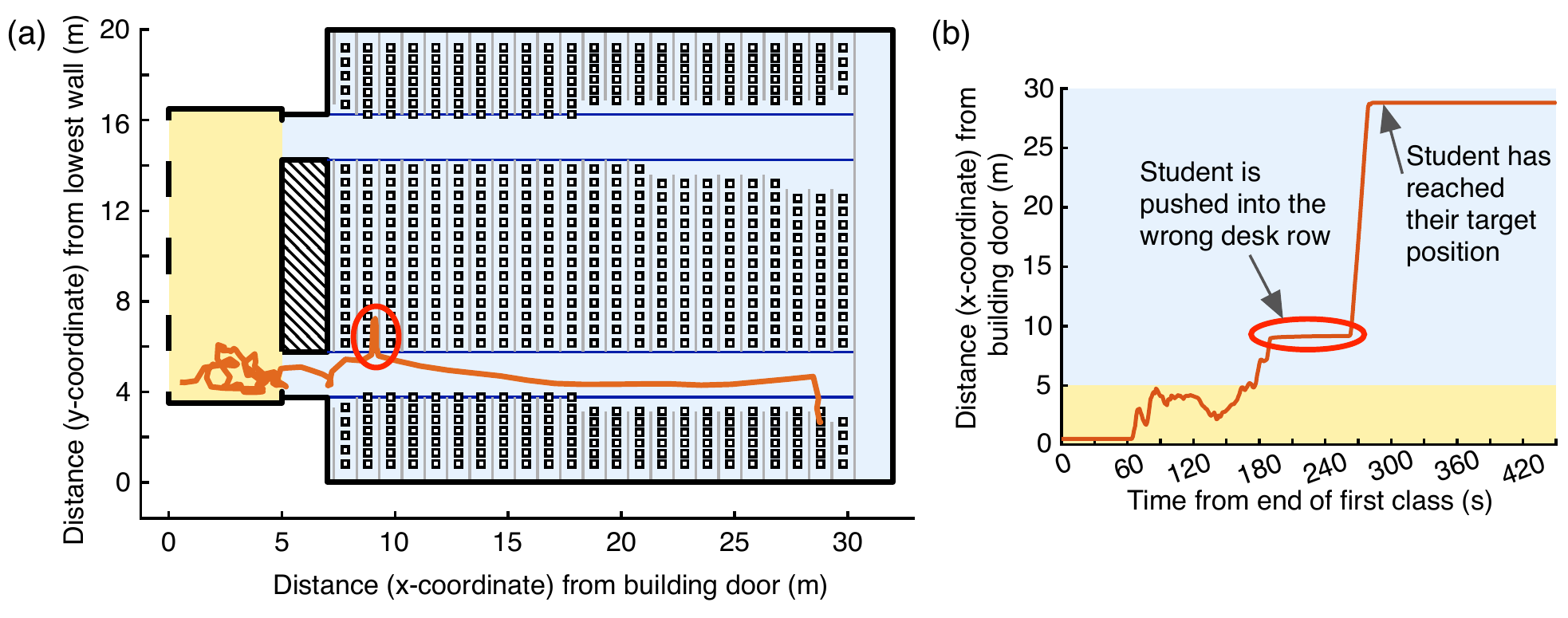}
    \caption{Example of an entering student with an outlier travel time, for the case of $600$ entering and $600$ exiting students in our $600$-person room with a separation time of $60$~s. We define ``outlier" as an agent with a travel time above $\text{Q}_3 + 1.5(\text{Q}_3 - \text{Q}_1)$ or below $\text{Q}_1 - 1.5(\text{Q}_3 - \text{Q}_1)$, where we compute the $25$th percentile $\text{Q}_1$ and $75$th percentile $\text{Q}_3$ across our $100$ simulations; also see Figure~7(c) in the main manuscript. (a) Here we show the trajectory of an example outlier agent, and we omit plotting the other agents in the simulation to make the trajectory visible. The example student attempts to enter the classroom, but is pushed back in the vestibule by the exiting class. Once the agent finally enters the classroom, they are pushed through our aisle-wall boundaries into the wrong row by other students. (b) To highlight the time that this student spends in different regions of the classroom, we also show the agent's $x$-coordinate in time. When the student is pushed into the wrong desk row, they are ``stuck" for about a minute, until the attractive forces that they feel toward their aisle destination overcome the repulsive forces that they feel from the aisle-wall boundaries and other students. This dynamic is rare in our simulations and may result from high student density; it leads to a travel time that falls outside of the gray bars in Figure~7(c) in the main manuscript, representing an outlier.}
    \label{fig:exampleStuck}
\end{figure*}

The outlier highlighted in Supplementary Figure~\ref{fig:exampleStuck} has a travel time that falls outside of the range covered by the gray bars in Figure~7(c) in the main manuscript. Because our model is a simplification that does not include many real-life dynamics that may prolong classroom turnover in large lecture halls (e.g.,  pauses due to students talking to one another or rare events like individuals from the prior class forgetting something in the classroom and needing to come back to look for it), we allow these rare outliers in our simulations; importantly, they do not affect the interpretation of our results. 

\begin{figure*}
    \centering
    \includegraphics[width=\textwidth]{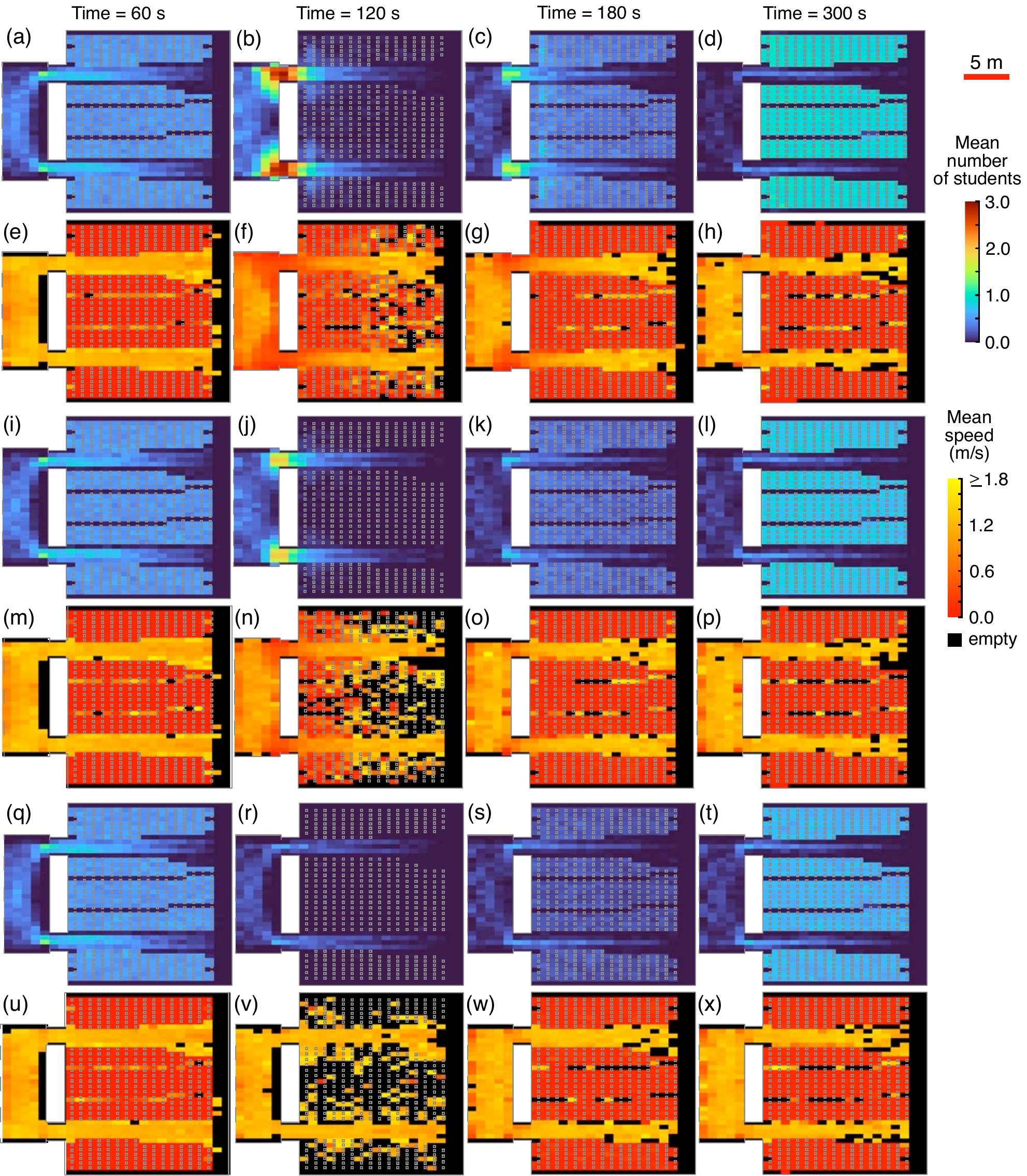}
    \caption{Mean student mass and speed across $100$ simulations of $400$ entering and $400$ exiting students in our Rock Hall, under different separation times. This figure expands on Figures~7(g)--(i) in the main manuscript. Mean mass (number of students in grid squares with area $1.5$~m$^2$) for a class-separation time of (a)--(d) $60$~s, (i)--(k) $90$~s, and (q)--(l) $120$~s across time. Similarly, (e)--(h), (m)--(n), and (u)--(x) show mean student speeds in grid squares at different times for a separation time of $60$~s, $90$~s, and $120$~s, respectively. We use the method described in Section~\ref{app:parameters} to compute the mean number of students in each grid square and the mean speed of students in these grid squares. Scale bar of $5$~m applies to all images, and times at the top of the figure apply to the six images in each column.}
    \label{fig:DelayBidirDensity}
\end{figure*}

In Supplementary Figure~\ref{fig:DelayBidirDensity}, we show the mean mass (number of students in grid squares with area $1.5$~m$^2$) and speed of students across $100$ simulations of $400$ entering and $400$ exiting students in our Rock Hall, under different separation times between when the first class ends and when entering students begin arriving. This figure expands on Figures~7(g)--(i) in the main manuscript, providing more timepoints. Crowding occurs near the classroom doors, and the congestion decreases as the separation time grows from $60$~s in Supplementary Figures~\ref{fig:DelayBidirDensity}(a)--(d) to $90$~s in Supplementary Figures~\ref{fig:DelayBidirDensity}(i)--(l) and $120$~s in Supplementary Figures~\ref{fig:DelayBidirDensity}(q)--(t). Crowding is accompanied by low speeds near classroom doors, and speeds in the vestibule grow as the separation time increases (e.g., compare Supplementary Figure~\ref{fig:DelayBidirDensity}(f) with a separation time of $60$~s to Supplementary Figure~\ref{fig:DelayBidirDensity}(v) with a separation time of $120$~s).

To facilitate further exploration into Figure~7 in the main manuscript,
Supplementary Tables \ref{tab:entrancetable} and \ref{tab:exittable} provide summary statistics for each of our sets of $100$ simulations for various class sizes and simulation conditions. The scenarios that we consider are pedestrians entering or exiting the classroom in the absence of social forces; a single entering class without an exiting class; a single exiting class without an entering class; 
and cases of bidirectional movement with two consecutive classes under separation times of $60$ s, $90$ s, or $120$~s between when the first class ends and when students from the next class begin arriving at the building doors. We present the mean, median, $75$th percentile, and $90$th percentile of the travel times in Supplementary Tables \ref{tab:entrancetable} and \ref{tab:exittable} for entering and exiting students, respectively, showing how these quantities vary for class sizes ranging from $200$ to $600$ students. 

\begin{table}
\caption{Summary statistics for student entrance times across $100$ simulations for various scenarios and class sizes. We present the (a) mean, (b) median, (c) $75$th percentile, and (d) $90$th percentile travel times for all of the entering students across our $100$ simulations; also see Figure~7 in the main manuscript and Supplementary Table II. ``No Social" refers to setting $B^\text{rep}=B^\text{col}=0$ in Equation~(9) in the main manuscript, so that students experience no forces from each other; this means that, from each student's perspective, they are the only student in the classroom. ``Enter Only" refers to the case of an entering class in the absence of an exiting class, and the cases of bidirectional movement with separation times of $60$ s, $90$ s, and $120$~s are labelled ``60 s Delay", ``90~s Delay", and ``120 s Delay", respectively.}
\label{tab:entrancetable}
\centering
\begin{subtable}{\textwidth}
\centering
\begin{tabular}{p{2cm} p{1.1cm} p{1.1cm} p{1.2cm} p{1cm} p{0.9cm}}
\textbf{Classroom}  & \multicolumn{5}{c}{\textbf{Class Size}} \\
\textbf{Scenario} & \textbf{200} & \textbf{300} & \textbf{400}  & \textbf{500}  & \textbf{600} \\
\hline
No Social & 13.73  &	15.66 &	16.76  &	17.84 & 19.33 \\
Enter Only & 14.61  &	16.62 & 	18.03 &	19.51 &	21.28 \\
120 s Delay & 14.65 & 16.73 &	18.25 & 19.80 &	22.35  \\
90 s Delay &  15.52 &	18.74 &	22.075 &	27.09 &	34.84  \\
60 s Delay &  17.54 &	24.65 &	33.09 &	43.65 &	55.83 \\
\hline
\end{tabular}
\caption{Mean travel time for entering students in seconds}
\end{subtable}
\centering
\begin{subtable}{\textwidth}
\centering
\begin{tabular}{p{2cm} p{1.1cm} p{1.1cm} p{1.2cm} p{1cm} p{0.9cm}}
\textbf{Classroom}  & \multicolumn{5}{c}{\textbf{Class Size}} \\
\textbf{Scenario} & \textbf{200} & \textbf{300} & \textbf{400}  & \textbf{500}  & \textbf{600} \\
\hline
No Social & 13.45 &	15.29 &	16.36 &	17.41 &	18.85\\
Enter Only & 14.24 &	16.24 &	17.5 &	18.81 &	20.45 \\
120 s Delay & 14.28 &	16.33	 & 17.75 &	 19.09 &	21.00 \\
90 s Delay & 14.87 &	17.45 &	19.35 &	21.89 &	25.92 \\
60 s Delay & 15.79 &	19.39 &	23.43 &	29.13 &	36.57\\
\hline
\end{tabular}
\caption{Median travel time for entering students in seconds}
\end{subtable}
\begin{subtable}{\textwidth}
\centering
\begin{tabular}{p{2cm} p{1.1cm} p{1.1cm} p{1.2cm} p{1cm} p{0.9cm}}
\textbf{Classroom}  & \multicolumn{5}{c}{\textbf{Class Size}} \\
\textbf{Scenario} & \textbf{200} & \textbf{300} & \textbf{400}  & \textbf{500}  & \textbf{600} \\
\hline
No Social& 16.10 & 18.59 & 20.00 & 21.44 & 23.60 \\
Enter Only & 17.05 & 19.52 & 21.21 & 22.91 & 25.21 \\
120 s Delay & 17.08	& 19.68 & 21.47	& 23.24 & 25.96 \\
90 s Delay & 18.01 &	21.38 &24.37 &	28.31 &	34.91\\
60 s Delay & 19.74 & 26.67 &36.96 & 48.89 & 62.06 \\
\hline
\end{tabular}
\caption{$75$th percentile entrance time in seconds}
\end{subtable}

\begin{subtable}{\textwidth}
\centering
\begin{tabular}{p{2cm} p{1.1cm} p{1.1cm} p{1.2cm} p{1cm} p{0.9cm}}
\textbf{Classroom}  & \multicolumn{5}{c}{\textbf{Class Size}} \\
\textbf{Scenario} & \textbf{200} & \textbf{300} & \textbf{400}  & \textbf{500}  & \textbf{600} \\
\hline
No Social & 18.63 &	21.65 &	 23.45 & 25.29 & 27.9 \\
Enter Only & 19.72 & 22.63 & 24.75 & 26.745 & 29.60 \\
120 s Delay & 19.81 & 22.8 & 25.005 &	27.17 & 30.555 \\
90 s Delay& 21.38 & 26.12 & 31.73 & 39.85 & 54.53 \\
60 s Delay & 25.87 & 42.01 & 62.235 & 84.67 & 113.70 \\
\hline
\end{tabular}
\caption{$90$th percentile entrance time in seconds}
\end{subtable}
\end{table}

\begin{table}
\caption{Summary statistics for student exit times across $100$ simulations for various scenarios and class sizes. We present the (a) mean, (b) median, (c) $75$th percentile, and (d) $90$th percentile travel times for all of the exiting students across our $100$ simulations; also see Figure~7 in the main manuscript and Supplementary Table I. ``No Social" refers to setting $B^\text{rep} =B^\text{col} =0$ in Equation~(9) in the main manuscript, so that students experience no forces from each other; this is equivalent to a single student exiting the classroom. The case of an exiting class in the absence of an entering class  is labelled ``Exit Only". The cases of bidirectional movement with separation times of $60$ s, $90$ s, and $120$ s between when the first class ends and when students from the next class begin arriving are labelled ``60 s Delay", ``90~s Delay", and ``120 s Delay", respectively.}
\label{tab:exittable}
\centering
\begin{subtable}{\textwidth}
\centering
\begin{tabular}{p{2cm} p{1.1cm} p{1.1cm} p{1.2cm} p{1cm} p{0.9cm}}
\textbf{Classroom}  & \multicolumn{5}{c}{\textbf{Class Size}} \\
\textbf{Scenario} & \textbf{200} & \textbf{300} & \textbf{400}  & \textbf{500}  & \textbf{600} \\
\hline
No Social & 73.54 & 75.43 & 76.54 & 77.73 & 79.18  \\
Exit Only & 74.91 & 77.35 & 78.54 & 80.11 & 82.02  \\
120 s Delay & 75.21 & 77.14 & 78.90 & 80.76 & 82.74 \\
90 s Delay & 76.25 & 78.54 & 80.64 & 83.99 & 87.43 \\
60 s Delay & 77.33 & 82.17 & 86.89 & 92.21 & 97.23\\
\hline
\end{tabular}
\caption{Mean travel time for exiting students in seconds}
\end{subtable}
\centering
\begin{subtable}{\textwidth}
\centering
\begin{tabular}{p{2cm} p{1.1cm} p{1.1cm} p{1.2cm} p{1cm} p{0.9cm}}
\textbf{Classroom}  & \multicolumn{5}{c}{\textbf{Class Size}} \\
\textbf{Scenario} & \textbf{200} & \textbf{300} & \textbf{400}  & \textbf{500}  & \textbf{600} \\
\hline
No Social & 73.685 & 75.34 & 76.42 & 77.76 & 79.2 \\
Exit Only & 75.08 & 77.99 & 78.97 & 80.78 & 82.795 \\
120 s Delay& 75.575 & 77.62 & 79.37 & 81.07 & 82.83  \\
90 s Delay & 75.925 & 77.28 & 78.63 & 80.925 & 82.45  \\
60 s Delay& 76.09 & 77.62 & 79.37 & 81.07 & 87.26 \\
\hline
\end{tabular}
\caption{Median travel time for exiting students in seconds}
\end{subtable}
\begin{subtable}{\textwidth}
\centering
\begin{tabular}{p{2cm} p{1.1cm} p{1.1cm} p{1.2cm} p{1cm} p{0.9cm}}
\textbf{Classroom}  & \multicolumn{5}{c}{\textbf{Class Size}} \\
\textbf{Scenario} & \textbf{200} & \textbf{300} & \textbf{400}  & \textbf{500}  & \textbf{600} \\
\hline
No Social& 95.155 & 96.94 & 97.8 & 99.18 & 100.97 \\
Exit Only & 96.305 & 98.98 & 100.48 & 102.15 & 104.32 \\
120 s Delay & 96.795 & 98.57 & 100.4 & 102.45 & 104.4  \\
90 s Delay & 96.955 & 99.21 & 101.24 & 104.02 & 106.55 \\
60 s Delay& 100.77 & 106.61 & 110.125 & 114.05 & 117.03  \\
\hline
\end{tabular}
\caption{$75$th percentile exit time in seconds}
\end{subtable}

\begin{subtable}{\textwidth}
\centering
\begin{tabular}{p{2cm} p{1.1cm} p{1.1cm} p{1.2cm} p{1cm} p{0.9cm}}
\textbf{Classroom}  & \multicolumn{5}{c}{\textbf{Class Size}} \\
\textbf{Scenario} & \textbf{200} & \textbf{300} & \textbf{400}  & \textbf{500}  & \textbf{600} \\
\hline
No Social& 112.325 & 114.385 & 115.515 & 117.01 & 118.46 \\
Exit Only & 113.65 & 116.00 & 117.185 & 118.89 & 120.855 \\
120 s Delay & 113.475 & 115.59 & 117.33 & 118.94 & 120.98 \\
90 s Delay& 116.66 & 120.99 & 123.59 & 127.225 & 131.42  \\
60 s Delay& 119.90 & 129.705 & 139.205 & 150.50 & 165.225  \\
\hline
\end{tabular}
\caption{$90$th percentile exit time in seconds}
\end{subtable}
\end{table}

\end{document}